\documentclass[reqno]{amsart}
\usepackage{oldlfont,amssymb}
\theoremstyle{plain}
\usepackage[OT2,OT1]{fontenc}
\newcommand\cyr{%
  \renewcommand\rmdefault{wncyr}%
  \renewcommand\sfdefault{wncyss}%
  \renewcommand\encodingdefault{OT2}%
  \normalfont
  \selectfont}
\DeclareTextFontCommand{\textcyr}{\cyr}
\newtheorem{lemma}{Lemma}[section]
\newtheorem{satz}{Proposition}[section]
\newtheorem{theorem}{Theorem}[section]
\newtheorem{corolla}{Corollary}[section]
\theoremstyle{definition}

\newtheorem{defi}{Definition}[section]

\theoremstyle{remark}

\newtheorem{remark}{Remark}[section]
\numberwithin{equation}{section}

\newcommand{\tr}{\mathop{\mathrm{tr}}}
\newcommand{\Tr}{\mathop{\mathrm{Tr}}}

\DeclareMathOperator*{\spec}{spec}
\begin{document}
\title[Connes' Trace Formula]{Connes' Trace Formula and Dirac Realization of Maxwell and Yang-Mills Action}
\thanks{Expanded version of a talk given at the workshop `The Standard Model of Elementary Particle Physics
from a Mathematical-Geometrical Viewpoint', held at the Ev.-Luth.~Volkshochschule Hesselberg near
Gerolfingen, Germany, March 14--19, 1999,
supported by the Volkswagenstiftung}
\author[P.M.\,Alberti, R.\,Matthes]{Peter M. Alberti \and Rainer Matthes}
\address{
Institute of Theoretical Physics\\
University of Leipzig\\
Augustusplatz 10, D-04109 Leipzig, Germany
}
\email{Peter.Alberti@itp.uni-leipzig.de \and Rainer.Matthes@itp.uni-leipzig.de}
\begin{abstract}
The paper covers known facts about the Dixmier trace (with some generalities about traces), the Wodzicki residue,
and Connes' trace theorem, including two variants of proof of the latter. Action formulas are treated very sketchy,
because they were considered in other lectures of the workshop.
\end{abstract}
\maketitle
\tableofcontents
\setcounter{section}{0}
\section{The Dixmier-Traces}\label{1}
The essential ingredients of the famous trace formula \cite{Conn:88,Conn:94}
of A.\,Connes are operator algebraic
constructs over the ${\mathsf W^*}$-algebra ${\mathsf B}({\mathcal H})$ of
all bounded linear operators over some infinite dimensional
separable Hilbert space ${\mathcal H}$ which are known as
{\em Dixmier-traces}.
The constructions will be explained in this section. For
a general operator-algebraic background the reader is referred e.g.~to
\cite{Dixm:64,Saka:71,Take:79,KaRi:83,KaRi:86}.
\subsection{Generalities on traces on ${\mathsf C}^*$- and ${\mathsf W}^*$-algebras}\label{1.1}
\subsubsection{Basic topological notions, notations}\label{1.1.1}
A ${\mathsf C}^*$-algebra $M$ is a Banach $^*$-algebra, with $^*$-operation
$x\,\longmapsto\,x^*$ and norm $\|\cdot\|$ obeying
$\|x^*x\|=\|x\|^2$, for each $x\in M$ ($^*$-quadratic property).
Let $M_+=\{x^*x : x\in M\}$
be the cone of positive elements of $M$. For $x\in M$ let $x\geq 0$ be synonymous
with $x\in M_+$. The Banach space of all continuous linear forms on $M$
(dual) will be denoted by $M^*$, the dual norm (functional norm)
be $\|\cdot\|_1$. As usual, a linear form $f$ over $M$ is termed positive,
$f\geq 0$, if $f(x^*x)\geq 0$, for each $x\in M$.
Remind that positive linear forms are automatically continuous and are
generating for $M^*$. Thus $M_+^*=\{f\in M^*:f\geq 0\}$ is the set of all these
forms. There is a fundamental result of operator theory saying that if
$M$ possesses a unit ${\mathbf 1}$ (unital ${\mathsf C}^*$-algebra),
then $f\geq 0$ if, and only if,
$f({\mathbf 1})=\|f\|_1$. It is common use to refer to
positive linear forms of norm one as states.
Thus, in a unital ${\mathsf C}^*$-algebra the set of all states on $M$,
${\mathcal S}(M)$, is easily seen to be a convex set which according to
the Alaoglu-Bourbaki theorem is $\sigma(M^*,M)$-compact. Here, the
$\sigma(M^*,M)$-topology (also $w^*$-topology in the special context at hand)
is the weakest locally convex topology generated by the seminorms $\rho_x$,
$x\in M$, with $\rho_x(f)=|f(x)|$, for each $f\in M^*$. The generalization
of the notion of positive linear form on $M$ (which refers to positive linear maps into
the special ${\mathsf C}^*$-algebra of complex numbers ${\mathbb C}$) is the notion of
the positive linear map. Say that a linear mapping $T:M\,\longrightarrow\, N$
which acts from one ${\mathsf C}^*$-algebra $M$ into another one $N$ is positive if $T(x^*x)\geq
{\mathbf 0}$ within $N$ for each $x\in M$. In the unital case (for $M$)
one then knows that $\|T({\mathbf 1})\|=\|T\|$ holds ($\|T\|$ refers to the
operator norm of $T$ as a linear operator acting from the one Banach space $M$
into the other $N$). On the other hand,
each linear map $T:M\,\longrightarrow\, N$ which obeys this relation is known to
be positive. If both $M$ and $N$ have a unit, a linear map $T:M\,\longrightarrow\, N$
is said to be unital if $T({\mathbf 1})={\mathbf 1}$ is fulfilled (the units being the
respective units). Thus, and in particular, each unital linear map
$T:M\,\longrightarrow\, N$ of norm one beween unital ${\mathsf C}^*$-algebras
has to be a positive map.

Remind that a ${\mathsf C}^*$-algebra $M$ is a ${\mathsf W}^*$-algebra, that is, is
$^*$-isomorphic to some $vN$-algebra on some Hilbert space ${\mathcal H}$, if and only if,
there exists a (unique) Banach subspace $M_*$ (the predual space) of $M^*$
such that $M$ is the (continuous) dual of $M_*$, $M=(M_*)^*$. Note that a
non-zero ${\mathsf W}^*$-algebra is always unital. Suppose now that
$M$ is a ${\mathsf W}^*$-algebra. The forms of $M_*$ will be referred to as normal
linear forms. One then knows that the normal positive linear forms
$M_{*+}=M_*\cap M_+^*$ (resp.~the normal states ${\mathcal S}_0(M)=
M_*\cap {\mathcal S}(M)$) are generating for $M_*$ in the sense that each
normal linear form may be represented as a complex linear combination of at most
four normal states.
As a consequence of this,
for each ascendingly directed (in the sense of $\leq$) bounded net
$\{x_\alpha\}\subset M_+$ there exists a lowest upper bound
$\mathop{l.u.b.} x_\alpha$ within $M_+$. On the other hand, a positive linear form
$\omega\in M_+^*$ is normal if, and only if, for each ascendingly directed bounded net
$\{x_\alpha\}\subset M_+$ the relation $\omega(\mathop{l.u.b.} x_\alpha)=
\mathop{l.u.b.} \omega(x_\alpha)=\lim_\alpha \omega(x_\alpha)$ is valid.
Note that in the latter characterization it suffices if the mentioned continuity
be fulfilled for ascendingly directed nets of orthoprojections of $M$.

In contrast to the previous,
$\nu\in M_+^*$ is called singular
if to each orthoprojection
$p\in M$ with $\nu(p)>0$ there is another orthoprojection $q\in M$ with
${\mathbf 0}<q<p$ and $\nu(q)=0$. According to \cite{take:58} each
$\omega\in M_+^*\backslash M_{+*}$ in a unique way can be decomposed as
$\omega=\omega_1 + \omega_2$, with normal $\omega_1\in M_{*+}$ and
singular $\omega_2\in M_+^*$.

The simplest example of a ${\mathsf W^*}$-algebra where singular positive linear forms can exist is
the commutative ${\mathsf W^*}$-algebra $\ell^\infty=\ell^\infty(\mathbb{N})$
of all bounded sequences $x=(x_n)=(x_1,x_2,\ldots)$
of complex numbers,
with norm $\|x\|_\infty=\sup_{n\in {\mathbb N}} |x_n|$, and algebra
multiplication $x\cdot y=(x_n y_n)$ and $^*$-operation $x^*=(\bar{x}_n)$
defined componentwise. Recall that in this case the predual space $\ell_*^\infty$ is the
Banach space of all absolutely summable
sequences $\ell^1(\mathbb{N})$, with norm
$\|\omega\|_1=\sum_{n\in {\mathbb N}} |\omega_n|$ for $\omega=(\omega_n)
\in \ell^1({\mathbb N})(=\ell^1)$.
Thereby, each such $\omega$ can be identified with an element in the dual Banach
space $(\ell^\infty)^*$ via the identification with the linear functional $\omega(\cdot)$ given as
$\omega(x)=\sum_{n\in {\mathbb N}} \omega_n x_n$, for each $x\in \ell^\infty$. For simplicity, also this
functional $\omega(\cdot)$ will be referred to as $\omega$, $\omega=\omega(\cdot)$.

In generalizing from the setting of a normal positive linear form, call a positive linear map
$T:M\,\longrightarrow\, N$ from one ${\mathsf W^*}$-algebra $M$ into another ${\mathsf W}^*$-algebra
$N$ normal if $T(\mathop{l.u.b.} x_\alpha)=
\mathop{l.u.b.} T(x_\alpha)$ holds for each ascendingly directed bounded net $\{x_\alpha\}\subset M_+$.
Note that in a ${\mathsf W}^*$-algebra the
Alaoglu-Bourbaki theorem may be applied on $M$, and then yields that the
(closed) unit ball $M_1$ of $M$ is $\sigma(M,M_*)$-compact. From this it
follows that the unit ball within the bounded linear operators which map the
${\mathsf W}^*$-algebra $M$ into itself is compact with respect to the
topology determined by the system of seminorms $\rho_{x,f}$, labelled by
$x\in M$ and $f\in M_*$, and which are defined at $T$ by
$\rho_{x,f}(T)=|f\circ T(x)|$. Refer to this topology as the
$\sigma(M,M_*)$-weak operator topology on the Banach algebra of bounded linear operators
${\mathsf B}(M)$ over the Banach space $M$. Thereby, by convention
for $T,S\in {\mathsf B}(M)$ let the product $TS\in {\mathsf B}(M)$ be defined
through successive application of maps to the elements of $M$ in accordance with the rule `apply right factor first',
that is $TS(x)=(T\circ S)(x)=T(S(x))$.
\subsubsection{Traces on ${\mathsf C}^*$- and ${\mathsf W}^*$-algebras}\label{1.1.2}
We recall the very basic facts on traces as found e.g.~in
\cite[6.1.]{Dixm:64}. A function
$\tau:M_+\,\longrightarrow\,\overline{{\mathbb R}}_+$ from the positive cone
into the extended positive reals is said to be a {\em trace} provided it is
(extended) {\em additive}, {\em positive homogeneous}\,(that is, $x,y\in M_+$
and $\lambda\in {{\mathbb R}}_+$ imply
$\tau(x+y)=\tau(x)+\tau(y)$ and $\tau(\lambda x)=\lambda \tau(x)$,
with $0\cdot\infty=0$ by convention)
and obeys $\tau(x^*x)=\tau(xx^*)$, for each $x\in M$ ({\em invariance property}).
>From the first two properties it follows that
$M_{\tau+}=\{x\in M_+:\tau(x)<\infty\}$ is a {\em hereditary
subcone} of $M_+$, that
is, $M_{\tau+}$ is a cone such that $x\in M_{\tau+}$ and
${\mathbf 0}\leq y\leq x$ for $y\in M_+$ implies $y\in M_{\tau+}$. From this it
is then easily inferred that ${\mathcal L}_\tau=\{x\in M:\tau(x^*x)<\infty\}$
is a {\em left ideal} in $M$. Owing to the invariance condition
$\tau(x^*x)=\tau(xx^*)$ however, ${\mathcal L}_\tau$ has to be also a right ideal.
Hence, ${\mathcal L}_\tau$ is a {\em two-sided ideal} of $M$, which is
characteristic for the trace $\tau$.
It is known that the complex linear span $[M_{\tau+}]$ is a $^*$-subalgebra of
$M$ which is even a two-sided ideal and which
is generated by ${\mathcal L}_\tau$ as $[M_{\tau+}]={\mathcal L}_\tau^2$.
Also, on the $^*$-subalgebra $[M_{\tau+}]$ there exists a unique (complex) linear form
$\tilde{\tau}$ obeying $\tilde{\tau}(x)=\tau(x)$, for each $x\in M_{\tau+}$. Owing to
invariance then also the characteristic commutation property
$\tilde{\tau}(yx)=\tilde{\tau}(xy)$ holds, for each $x\in [M_{\tau+}]$ and all $y\in M$.

The two-sided ideal $[M_{\tau+}]$ will be referred to as
{\em defining ideal}\, of the trace $\tau$. Since the linear extension $\tilde{\tau}$ of
$\tau$ from the cone of all positive elements of $[M_{\tau+}]$
(which according to the previous is $M_{\tau+}$) onto
$[M_{\tau+}]$ is unique, by tacit understanding the notation $\tau(x)$ will be also used at
non-positive $x$ of the defining ideal of $\tau$ if the evaluation $\tilde{\tau}(x)$ of the linear functional
$\tilde{\tau}$ at $x$ is meant.

The trace $\tau$ is termed {\em finite trace} if $M_{\tau+}=M_+$, and
{\em semifinite trace} if
$\tau(x)=\sup \{\tau(y):y\leq x,\,y\in M_{\tau+}\}$, for each $x\in M_+$. If $M$ is a ${\mathsf W^*}$-algebra,
with group of unitary elements ${\mathcal U}(M)$, the above condition on invariance usually is
replaced with a seemingly weaker requirement upon {\em unitary invariance},
that is $\tau(u^*x u)=\tau(x)$ be fulfilled,
for each $x\in M_+$ and $u\in {\mathcal U}(M)$. However, both conditions are equivalent there
(and are so even on unital ${\mathsf C}^*$-algebras). Also, in the ${\mathsf W^*}$-case the trace
$\tau$ is said to be
{\em normal} provided for each ascendingly directed bounded net
$\{x_\alpha\}\subset M_+$ the relation $\tau(\mathop{l.u.b.} x_\alpha)=
\mathop{l.u.b.} \tau(x_\alpha)=\lim_\alpha \tau(x_\alpha)$ is fulfilled.\\
Now, suppose $I\subset M$ is a proper two-sided ideal of the ${\mathsf W^*}$-algebra
$M$. Then, $I$ is also a $^*$-subalgebra of $M$, with generating positive cone
$I_+=I\cap M_+$, that is, $I=[I_+]$ is fulfilled
(these facts are consequences of the polar decomposition theorem,
essentially). Under these premises we have the following extension principle\,:
\begin{lemma}\label{nullaux}
Suppose $I_+$ is a hereditary subcone of $M_+$. Then, each additive, positive homogeneous
and invariant map $\tau_0:I_+\,\longrightarrow\,\overline{{\mathbb R}}_+$ extends
to a trace $\tau$ on $M$, with $M_{\tau+}=\{x\in I_+:\tau_0(x)<\infty\}$.
\end{lemma}
\begin{proof}
Define $\tau(x)=\tau_0(x)$, for $x\in I_+$, and $\tau(x)=\infty$
for $x\in M_+\backslash I_+$. Then, since $I_+$ is a hereditary cone,
for $x,y\in M_+$ with $x+y\in I_+$ one has both $x,y\in I_+$, and thus
$\tau(x)+\tau(y)=\tau(x+y)$ is evident from
$\tau_0(x)+\tau_0(y)=\tau_0(x+y)$. For $x,y\in M_+$ with
$x+y\not\in I_+$ at least one of $x,y$ must not be in $I_+$. Hence,
$\tau(x)+\tau(y)=\infty$ and $\tau(x+y)=\infty$ by definition of $\tau$.
Thus additivity holds in any case. That $\tau$ is also positive homogeneous
is clear. Finally, remind that, according to polar decomposition $x=u|x|$
for $x\in M$, one has $xx^*=u(x^*x)u^*$ and $x^*x=u^*(xx^*)u$, with the
partial isometry $u\in M$. Since $I$ is a two-sided ideal of $M$, from this
one infers $xx^*\in I_+$ if, and only if, $x^*x\in I_+$. From this in view of
the definition and since $\tau_0$ is invariant on $I_+$ also invariance of
$\tau$ on $M_+$ follows.
\end{proof}
\subsection{Examples of traces}\label{1.2}
In the following the classical special cases of traces on $M$ with either $M={\mathsf B}({\mathcal H})$
or $M={\mathsf {\mathcal C}B}({\mathcal H})$ are considered in more detail,
where ${\mathsf {\mathcal C}B}({\mathcal H})$
is the ${\mathsf C}^*$-subalgebra of ${\mathsf B}({\mathcal H})$ of all {\em compact} linear operators on the
separable infinite dimensional Hilbert space ${\mathcal H}$. For generalities on the theory of
compact operators and proofs from there the reader is referred to
\cite{Neum:59,Scha:70} e.g.;
in the following recall only those few facts and details which are important in the context of traces.
In all that follows,
the scalar product ${\mathcal H}\times {\mathcal H}
\ni\{\chi,\eta\}\,\longmapsto\,\langle\chi,\eta\rangle\in
{\mathbb{C}}$ on ${\mathcal H}$ by convention is supposed
to be linear with respect to the first argument $\chi$, and antilinear
in the second argument $\eta$, and maps into the complex
field ${\mathbb{C}}$.
\subsubsection{Traces on compact linear operators}\label{1.2.1}
We start with recalling the characterization of positive compact linear
operators in terms of a spectral theorem.
Let $x\in {\mathsf B}({\mathcal H})_+$ be a non-trivial
positive (=non-negative) bounded linear operator. Then,
$x$ is a (positive) compact operator,
$x\in {\mathsf {\mathcal C}B}({\mathcal H})_+$ if, and only if, the following
two condition are fulfilled. Firstly,
there have to exist a non-increasing infinite sequence
$(\mu_1(x),\mu_2(x),\ldots)$ of non-negative reals $\mu_k(x)$, which converge to zero as
$k\to\infty$, and an infinite
{\em orthonormal system} (o.n.s.~for short)
$\{\varphi_n\}\subset {\mathcal H}$ of
eigenvectors of
$x$ obeying $x\varphi_k=\mu_k(x)\varphi_k$, for each $k\in {\mathbb N}$, and
with $x\varphi=0$, for each $\varphi\in [\{\varphi_n\}]^\perp$ (thus the spectrum of $x$
is $\spec(x)=\{\mu_k(x): k\in {\mathbb N}\}\cup \{0\}$). And secondly,
each non-zero eigenvalue of $x$ has only
finite multiplicity, that is, $m(\mu)=\#\{k:\mu_k(x)=\mu\}$ obeys $m(\mu)<\infty$, for each $\mu\in
{\mathbb R}_+\backslash\{0\}$.

Recall that ${\mathsf {\mathcal C}B}({\mathcal H})$ is also a closed $^*$-ideal of
${\mathsf B}({\mathcal H})$. Hence, according to polar decomposition,
$x\in {\mathsf B}({\mathcal H})$ is compact if, and only if, the module $|x|=\sqrt{x^*x}$ of $x$ is
compact, $|x|\in {\mathsf {\mathcal C}B}({\mathcal H})_+$.
In line with this and following some common use, for
$x\in {\mathsf {\mathcal C}B}({\mathcal H})$ and in view of the above define $\mu_k(x)=\mu_k(|x|)$, for each
$k\in {\mathbb N}$, and
refer to the ordered sequence
$\mu_1(x)\geq \mu_2(x)\geq\ldots $ of eigenvalues of $|x|$ (with each of the
non-zero eigenvalues repeated according to its
multiplicity) as {\em characteristic sequence} of $x$. The terms of this sequence
can be obtained by
minimizing the distance of the given compact operator $x$ to the
finite rank linear operators (which are special compact operators) of a fixed rank as follows\,:
\begin{subequations}\label{sing}
\begin{equation}\label{sing.1}
\forall\,k\in {\mathbb N}:\ \mu_k(x)=\min\{\|x-y\|: y\in {\mathsf {\mathcal C}B}({\mathcal H}),\,\dim y{\mathcal H}\leq k\}\,.
\end{equation}
Alternatively, and yet more important, these values can be obtained
also from a representation of
the sequence $\{\sigma_n(x)\}$ of their partial sums
$\sigma_k(x)=\sum_{j\leq k}\mu_k(x)$ which arises
from maximizing the following expression over the unitaries ${\mathsf U}({\mathcal H})$ of ${\mathcal H}$
and finite orthonormal systems $\{\psi_1,\ldots,\psi_k\}\subset
{\mathcal H}$ of
cardinality $k\in {\mathbb N}$\,:
\begin{equation}\label{sing.2}
\sigma_k(x)=\max\biggl\{\biggl|\sum_{j\leq k}\langle u x\psi_j,\psi_j\rangle
\biggr|: u\in {\mathsf U}({\mathcal H}),\,\{\psi_1,\ldots,\psi_k\}\text{ o.n.s.}\biggr\}\,,
\end{equation}
which for positive $x$ simplifies into
\begin{equation}\label{sing.2a}
\forall\,x\in {\mathsf {\mathcal C}B}({\mathcal H})_+:\ \sigma_k(x)=\max\biggl\{\sum_{j\leq k}\langle x\psi_j,\psi_j\rangle
:\,\{\psi_1,\ldots,\psi_k\}\text{ o.n.s.}\biggr\}\,.
\end{equation}
Now, let us fix an arbitrary {\em maximal orthonormal system} (m.o.n.s.~for short)
$\{\varphi_n\}\subset {\mathcal H}$, and let $p$ be an
orthoprojection with $\dim p{\mathcal H}=k<\infty$. Then, for each
$x\in {\mathsf B}({\mathcal H})$ the operator $x p$ is of finite rank, and
for each
o.n.s.~$\{\psi_1,\ldots,\psi_k\}$
which linearily spans $p{\mathcal H}$, by elementary Hilbert space calculus
one derives the relation
\begin{equation}\tag{$\times$}
\sum_{n=1}^\infty\langle x p\varphi_n,\varphi_n\rangle=\sum_{j\leq k}\langle x
\psi_j,\psi_j\rangle\,.
\end{equation}
Hence, in case of compact $x$ (\ref{sing.2}) equivalently reads as
\begin{equation}\tag{$\star$}
\sigma_k(x)=\max\biggl\{\biggl|\sum_{n=1}^\infty\langle u x p\varphi_n,\varphi_n
\rangle\biggr|: u\in {\mathsf U}({\mathcal H}),\,p=p^*=p^2,
\dim p{\mathcal H}\leq k \biggr\}\,.
\end{equation}
Note that for $x\geq {\mathbf 0}$ the expression of ($\times$) is
positive and with the help of similarly elementary calculations as those which led to
($\times$) one infers that for each orthoprojection
$p$ with $\dim p{\mathcal H}=k<\infty$ and any $x\in{\mathsf B}({\mathcal H})_+$ the following holds\,:
\begin{equation}\tag{$\star\star$}
\sum_{j\leq k}\langle x
\psi_j,\psi_j\rangle=\sum_{n=1}^\infty \langle  p \sqrt{x}\varphi_n,\sqrt{x}
\varphi_n\rangle=\sum_{n=1}^\infty\langle x p\varphi_n,\varphi_n\rangle\geq 0\,.
\end{equation}
Especially, since according to
\eqref{sing.2a} for {\em positive} compact $x$ maximizing over the unitaries
becomes redundant and may be omitted,
in view of this and
({$\star\star$}) for each such element the relation
({$\star$}) then simplifies into the following well-known form\,:
\begin{equation}\label{sing.3}
\forall\,x\in {\mathsf {\mathcal C}B}({\mathcal H})_+:\,\sigma_k(x)=\max\biggl\{\sum_{n=1}^\infty\langle x p\varphi_n,\varphi_n
\rangle: p=p^*=p^2,
\dim p{\mathcal H}\leq k \biggr\}\,.
\end{equation}
\end{subequations}
>From \eqref{sing.2}--\eqref{sing.3} one now concludes
some useful relations and estimates. The first is a rather trivial consequence of the definition of $\sigma_k(x)$
and says that
\begin{subequations}\label{basis}
\begin{equation}\label{basis.1}
\forall\,x\in {\mathsf {\mathcal C}B}({\mathcal H}),\,\lambda\in
{\mathbb C},\,k\in {\mathbb N}:\
\sigma_k(\lambda x)=|\lambda|\,\sigma_k(x)\,.
\end{equation}
It is stated here only for completeness.
In the special case of positive compact
operators from \eqref{sing.2a} we get the following often used estimates\,:
\begin{equation}\label{basis.1a}
\forall\,x,y\in {\mathsf {\mathcal C}B}({\mathcal H})_+,\,k\in {\mathbb N}:\
y\leq x\,\Longrightarrow\,\sigma_k(y)\leq \sigma_k(x)\,.
\end{equation}
The next estimate is due to \cite{KyFa:51} and at once gets
obvious from \eqref{sing.2}, and tells us that the following holds\,:
\begin{equation}\label{basis.1b}
\forall\,x,y\in {\mathsf {\mathcal C}B}({\mathcal H}),\,k\in {\mathbb N}:\
\sigma_k(x+y)\leq \sigma_k(x)+\sigma_k(y)\,.
\end{equation}
The third estimate deals with an upper bound of $\sigma_k(x)+\sigma_k(y)$
in case of positive operators $x,y\in {\mathsf {\mathcal C}B}({\mathcal H})_+$
and arises from \eqref{sing.3}. In line with the latter, let orthoprojections
$p,q$ of rank $k$ be given such that $\sigma_k(x)=\sum_{n=1}^\infty\langle x p\varphi_n,\varphi_n
\rangle$ and $\sigma_k(y)=\sum_{n=1}^\infty\langle y q\varphi_n,\varphi_n
\rangle$ are fulfilled. Then, the least orthoprojection $p\vee q$ majorizing
both $p$ and $q$ has rank $2k$ at most. Thus there is an orthoprojection $Q$ of rank
$2 k$ and obeying $p\vee q\leq Q$. Hence, in view of the choice of $p,q$ and
with the help of ({$\star\star$}) one infers that $\sum_{n=1}^\infty\langle x Q\varphi_n,\varphi_n
\rangle=\sigma_k(x)+\sum_{n=1}^\infty\langle x(Q-p)\varphi_n,\varphi_n
\rangle\geq \sigma_k(x)$ and $\sum_{n=1}^\infty\langle y Q\varphi_n,\varphi_n
\rangle=\sigma_k(y)+\sum_{n=1}^\infty\langle x(Q-q)\varphi_n,\varphi_n
\rangle\geq \sigma_k(y)$. In view of \eqref{sing.3} from this then
$\sigma_k(x)+\sigma_k(y)\leq \sigma_{2k}(x+y)$ follows.
For positive compact operators the previous together with \eqref{basis.1b}
may be summarized into the following one\,:
\begin{equation}\label{basis.2}
\forall\,x,y\in {\mathsf {\mathcal C}B}({\mathcal H})_+,\,k\in {\mathbb N}:\
\sigma_k(x+y)\leq \sigma_k(x)+\sigma_k(y)\leq \sigma_{2k}(x+y)\,.
\end{equation}
Note that, since ${\mathsf {\mathcal C}B}({\mathcal H})$ is a two-sided ideal, from
\eqref{sing.1} for each $y\in
{\mathsf {\mathcal C}B}({\mathcal H})$ and $a,b\in {\mathsf B}({\mathcal H})$ the estimate
\begin{equation}\label{basis.2b}
\forall\,k\in{\mathbb N}:\ \mu_k(ayb)\leq\|a\|\,\|b\|\,\mu_k(y)
\end{equation}
can be obtained. Thus, under these conditions one has
\begin{equation}\label{basis.2a}
\forall\,a,b\in {\mathsf B}({\mathcal H}),\,y\in
{\mathsf {\mathcal C}B}({\mathcal H}),\,k\in {\mathbb N} :\ \sigma_k(ayb)\leq\|a\|\,\|b\|\,\sigma_k(y)\,.
\end{equation}
Especially, if $x=u|x|$ is the polar decomposition of
$x\in {\mathsf {\mathcal C}B}({\mathcal H})$ within ${\mathsf B}({\mathcal H})$,
then with the partial isometry
$u\in {\mathsf B}({\mathcal H})$ one has both, $xx^*=u x^*x u^*$ and $x^*x=u^* xx^* u$.
In the special cases of \eqref{basis.2a} with $y=xx^*,\,a=u^*,\,b=u$ and
$y=x^*x,\,a=u,\,b=u^*$ we arrive at estimates which fit together into the following assertion\,:
\begin{equation}\label{basis.3}
\forall\,x\in {\mathsf {\mathcal C}B}({\mathcal H}),\,k\in {\mathbb N}:\
\sigma_k(x^*x)=\sigma_k(x x^*)\,.
\end{equation}
\end{subequations}
In the following, a trace $\tau$ is said to be {\em non-trivial}
if there is at least one
$x\geq {\mathbf 0}$ with $0<\tau(x)<\infty$.
The relations given in eqs.~\eqref{basis} are the key facts that
the theory of traces on both algebras
${\mathsf {\mathcal C}B}({\mathcal H})$ and ${\mathsf B}({\mathcal H})$ can
be based on.
\begin{lemma}\label{spur.1}
Let $\tr:{\mathsf {\mathcal C}B}({\mathcal H})_+\,\longmapsto\,\overline{{\mathbb R}}_+$ be defined
by $\tr x=\lim_{n\to\infty} \sigma_n(x)$, for each
$x\in {\mathsf {\mathcal C}B}({\mathcal H})_+$.
Then, $\tr$ is a non-trivial semifinite trace on
${\mathsf {\mathcal C}B}({\mathcal H})$. Moreover,
to each non-trivial trace $\tau$ which does not vanish identically on
the positive operators of finite rank
there exists unique $\lambda\in {\mathbb R}_+\backslash\{0\}$ such
that $\lambda\cdot \tau(x)\geq \tr x$ holds for all $x\in{\mathsf {\mathcal C}B}({\mathcal H})_+$, and
with
equality occuring at each $x$ of finite rank.
\end{lemma}
\begin{proof}
The sequence $\{ \sigma_n(x)\}$ is increasing,
for each $x\in {\mathsf {\mathcal C}B}({\mathcal H})_+$. Thus
$\tr x=\lim_{n\to\infty} \sigma_n(x)$ exists in the extended sense. Especially, from \eqref{basis.2}
in the limit then additivity of $\tr$ follows, whereas from \eqref{basis.1} and \eqref{basis.3}
homogeneity and invariance can be seen.
Thus, $\tr$ is a trace (see \ref{1.1.2}). By construction $0<\tr x<\infty$ for
each compact positive $x\not={\mathbf 0}$ of finite rank. Thus $\tr$ is non-trivial.
However, since ${\mathcal H}$ is infinite dimensional, $\tr x=\infty$ will occur for some
positive compact operators. To see that $\tr$ is semifinite
requires to prove that for
$x\in {\mathsf {\mathcal C}B}({\mathcal H})_+$ with $\tr x=\infty$ there existed a sequence $\{x_n\}\subset
{\mathsf {\mathcal C}B}({\mathcal H})_+$ with $x_n\leq x$ and $\tr x_n <\infty$ such that
$\lim_{n\to\infty}
\tr x_n=\infty$. Note that by definition of $\tr$, $\tr x=\infty$  implies that $x$ cannot be of
finite rank.
Hence, $x$ can be written as $x=\sum_{k=1}^\infty \mu_k(x)\,p_k$, with infinitely many
mutually orthogonal one-dimensional orthoprojections $p_k$ and all $\mu_k(x)\not=0$.
Clearly, for each $n\in {\mathbb N}$ the operators
$x_n=\sum_{k=1}^n \mu_k(x)\,p_k$ are of finite rank and obey
${\mathbf 0}\leq x_1\leq x_2\leq x_3\leq \ldots \leq x$. Also, owing to
$\sigma_k(x_n)=\sigma_n(x)$ for $k\geq n$, one has $\tr x_n=\sigma_n(x)$, and therefore
$\lim_{n\to\infty}
\tr x_n=\infty$ follows. Thus $\tr$ is semifinite.
Suppose $\tau$ is a non-trivial trace. Thus $0<\tau(y)<\infty$, for some positive
compact $y$. Suppose $\tau(x)>0$ for some
$x\geq {\mathbf 0}$ of finite rank. According to additivity and homogeneity
of $\tau$ there has to exist a one-dimensional subprojection $p$ of a
spectral orthoprojection
of $x$ with $\tau(p)>0$.  The same arguments for
$y$ ensure that $\tau(q)<\infty$, for some one-dimensional subprojection
$q$ of some spectral projection of $y$. But since $q=vv^*$ and $p=v^*v$,
with $v\in{\mathsf {\mathcal C}B}({\mathcal H})$, by invariance of
$\tau$ one has $\tau(q)=\tau(p)$.  Hence
$\infty>\tau(p)>0$, and $\tau(q)=\tau(p)$ for each one-dimensional
orthoprojection $q$.
Put $\lambda=\tau(p)^{-1}$. Then $\lambda\cdot\tau(q)=\tr q$, and thus
$\lambda\cdot\tau(x)=\tr x$ for each positive operator $x$ of
finite rank. Finally, if $x\in {\mathsf {\mathcal C}B}({\mathcal H})_+$
is not of finite rank, let
${\mathbf 0}\leq x_1\leq x_2\leq x_3\leq \ldots \leq x$ be the above
approximating sequence of $x$ by finite rank
operators $x_n$. Also in such case $\lim_{n\to\infty} \tr x_n=\lim_{n\to\infty} \sigma_n(x)=\tr x$
follows. Hence, in view of the above relation over the operators of
finite rank, and since $\tau(x_n)\leq \tau(x)$ holds, $\tr x=\lim_{n\to\infty} \tr x_n=\lambda\,
\lim_{n\to\infty} \tau(x_n)\leq\lambda\cdot\tau(x)$.
\end{proof}
For completeness, we give yet the most famous formula relating $\tr$
and which makes that this trace is so extremely useful.
\begin{corolla}\label{spur}
For each maximal orthonormal system $\{\psi_n\}\subset {\mathcal H}$
and $x\in {\mathsf {\mathcal C}B}({\mathcal H})_+$
one has $\tr x=\sum_{n=1}^\infty \langle x\psi_n,\psi_n\rangle$.
\end{corolla}
\begin{proof}
Let $\{\varphi_j\}$ be an o.n.s.~with $x\varphi_k=\mu_k(x)\varphi_k$, for all $k\in {\mathbb N}$, and be
$p_n$ the orthoprojection with $p_n{\mathcal H}=[\varphi_1,\ldots,\varphi_n]$. Then, by positivity
of $x$ one has $\langle x\psi_n,\psi_n\rangle=\langle \sqrt{x}\psi_n,\sqrt{x}\psi_n\rangle\geq
\langle p_k \sqrt{x}\psi_n,\sqrt{x}\psi_n\rangle$, and therefore and in view of ($\star\star$) one
gets
$
\sum_{n=1}^\infty \langle x\psi_n,\psi_n\rangle\geq \sum_{n=1}^\infty
\langle p_k \sqrt{x}\psi_n,\sqrt{x}\psi_n\rangle=\sum_{j=1}^k \langle x\varphi_j,\varphi_j\rangle=
\sigma_k(x)$. According to Lemma \ref{spur.1} then
$\sum_{n=1}^\infty \langle x\psi_n,\psi_n\rangle\geq \tr x$ follows.
On the other hand, if $q_k$ is the orthoprojection onto
$[\psi_1,\ldots,\psi_k]$, according to
\eqref{sing.3} for each $k\in {\mathbb N}$ certainly
$\sum_{n=1}^k \langle x\psi_n,\psi_n\rangle=
\sum_{n=1}^\infty \langle x q_k\psi_n,\psi_n\rangle\leq \sigma_k(x)$. From this
in view of Lemma \ref{spur.1} once more again
$\sum_{n=1}^\infty \langle x\psi_n,\psi_n\rangle\leq \tr x$ is seen. Taking together this with the
above estimate provides that equality has to occur.
\end{proof}
A non-zero trace $\tau$ on ${\mathsf {\mathcal C}B}({\mathcal H})$ will be
said to be {\em singular}
if $\tau(x)=0$ for each $x\geq{\mathbf 0}$ of finite rank. Relating this and
non-trivial traces there is the following result.
\begin{corolla}\label{zerleg}
Let $\tau$ be a non-trivial trace on ${\mathsf {\mathcal C}B}({\mathcal H})$.
Then, either $\tau=\lambda\cdot\tr$ holds, for a unique $\lambda\in {\mathbb R}_+$,
or there exist a singular trace $\tau_s$ and a unique
$\alpha\in {\mathbb R}_+$ such that $\tau=\tau_s+\alpha\cdot\tr$.
\end{corolla}
\begin{proof}
If $\tau=\lambda\cdot\tr$ is fulfilled, then $\tau(p)=\lambda\,\tr p$, for each one-dimensional orthoprojection
$p$. Owing to $\tr p=1$ (see Corollary \ref{spur}) then $\lambda=\tau(p)$
follows.

Suppose $\tau\not\in{\mathbb R}_+\tr$. Then, $\tau\not=0$, and if a
decomposition $\tau=\tau_s+\alpha\cdot\tr$
with singular $\tau_s$ exists, then $\tau(p)=\alpha$,
for some (and thus any) one-dimensional orthoprojection $p$, and
the following two alternatives have to be dealt with\,:
firstly, if $\tau$ is vanishing on all positive operators of finite rank,
$\tau$ is singular, and $\tau=\tau_s$ and $\alpha=0$ have to be chosen
(see above). Secondly, if $\tau$ does not vanish on all positive operators
of finite rank, according to Lemma \ref{spur.1} there exists unique $\lambda>0$ with $\lambda\cdot\tau(x)
\geq \tr x$, for each $x\in{\mathsf {\mathcal C}B}({\mathcal H})_+$, with
equality occuring on any operator of finite rank. Hence, in defining
$\tau_s(x)=\tau(x)-\lambda^{-1}\tr x$, for each $x$ with $\tr x<\infty$, and
$\tau_s(x)=\infty$ else, we get a positive map $\tau_s$ which
does not vanish identically on the positive compact operators,
but which is vanishing on all positive operators of finite rank. From
the previous and since both $\tau$ and
$\tr$ are traces, also additivity, positive homogeneity and invariance of
$\tau_s$ at once follow.
Hence, $\tau_s$ is a singular trace, which is easily seen to obey
$\tau=\tau_s+\alpha\cdot \tr$, with $\alpha=\lambda^{-1}$.
\end{proof}
\subsubsection{Traces on ${\mathsf B}({\mathcal H})$}\label{1.2.2}
Remind in short the theory of traces on $M={\mathsf B}({\mathcal H})$,
with separable infinite dimensional Hilbert space ${\mathcal H}$.
Let ${\mathsf {\mathcal F}B}({\mathcal H})$ be the two-sided ideal of all operators of finit rank in
${\mathsf B}({\mathcal H})$. In the following an ideal ${\mathcal I}$ will be
termed {\em non-trivial} if ${\mathcal I}\not=\{{\mathbf 0}\}$ and
${\mathcal I}\not={\mathsf B}({\mathcal H})$.
Both ${\mathsf {\mathcal F}B}({\mathcal H})$ and
${\mathsf {\mathcal C}B}({\mathcal H})$ are non-trivial two-sided ideals.
Thereby, the compact operators form a closed ideal, with
${\mathsf {\mathcal F}B}({\mathcal H})$
being dense within ${\mathsf {\mathcal C}B}({\mathcal H})$.
Start with a useful criterion on non-compactness for a positive operator.
\begin{lemma}\label{noncomp}
A positive operator $x\geq {\mathbf 0}$ is non-compact if, and only if,
there exist real $\lambda>0$ and infinite dimensional orthoprojection
$p$ obeying $\lambda\,p\leq x$.
\end{lemma}
\begin{proof}
Note that, in contrast to the spectral characterization of positive compact
operators, the spectral theorem in case of a {\em non-compact}
$x\geq {\mathbf 0}$ with $\#\spec (x)<\infty$ provides that $\lambda\,p\leq x$
has to be fulfilled,
for some non-zero $\lambda$ and orthoprojection $p$ with
$\dim p{\mathcal H}=\infty$ (for one
$\lambda\in \spec (x)\backslash\{0\}$ at least the
corresponding spectral eigenprojection $p$ has to meet the requirement). But then,
due to normclosedness of the compact operators, and since for each
positive $x$ one has $x\in\overline{\{y: {\mathbf 0}\leq y\leq x,
\#\spec(x)<\infty\}}$ (uniform closure), such type of estimate has to
exist in each case of a non-compact positive operator $x$. On the other hand,
if $\lambda\,p\leq x$ is fulfilled,
for some non-zero $\lambda$ and infinite dimensional orthoprojection $p$,
in view of this relation the equivalence of $p$ with the unit
operator ${\mathbf 1}$ will imply $v^*x v$ to be invertible, for the
partial isometry $v$ achieving $p=vv^*$, ${\mathbf 1}=v^*v$. Thus,
owing to the non-triviality
of the ideal ${\mathsf {\mathcal C}B}({\mathcal H})$,
$v^*x v\not\in {\mathsf {\mathcal C}B}({\mathcal H})$ has to hold.
Due to two-sidedness
of ${\mathsf {\mathcal C}B}({\mathcal H})$ the latter requires that also
$x$ was non-compact.
\end{proof}
\begin{corolla}\label{here}
Both ${\mathsf {\mathcal F}B}({\mathcal H})_+$ and
${\mathsf {\mathcal C}B}({\mathcal H})_+$ are hereditary subcones of ${\mathsf B}({\mathcal H})_+$.
\end{corolla}
\begin{proof}
For ${\mathsf {\mathcal F}B}({\mathcal H})_+$ the assertion is trivial. For
non-zero $x\in {\mathsf {\mathcal C}B}({\mathcal H})_+$ and positive
$y\not={\mathbf 0}$ with $y\leq x$ also $y$ must be compact since otherwise
the criterion of Lemma \ref{noncomp} were applicable to $y$ with
resulting in a contradiction to the assumed
compactness of $x$, by the same criterion.
\end{proof}
The following is likely the most remarkable result relating ideals in
${\mathsf B}({\mathcal H})$ and descends from \cite{Calk:41}, see also \cite[Lemma 11,\,Theorem 11]{Scha:70}.
\begin{theorem}\label{calkin}
${\mathsf {\mathcal F}B}({\mathcal H})\subset {\mathcal I}\subset {\mathsf {\mathcal C}B}({\mathcal H})$,
for each non-trivial, two-sided ideal ${\mathcal I}$.
\end{theorem}
As a consequence of this the defining ideal of a non-trivial trace
$\tau$ on ${\mathsf B}({\mathcal H})$ always is a non-zero ideal of
compact operators. Thus especially $\tau(x)=\infty$ must be fulfilled,
for each $x\in {\mathsf B}({\mathcal H})_+\backslash {\mathsf {\mathcal C}B}({\mathcal H})_+$.
On the other hand, since according to Corollary \ref{here}
${\mathsf {\mathcal C}B}({\mathcal H})_+$ is a hereditary cone, whenever
$\tau_0$ is a non-zero trace on ${\mathsf {\mathcal C}B}({\mathcal H})$,
then the extension principle of
Lemma \ref{nullaux} can be applied and shows that
upon defining $\tau(x)=\tau_0(x)$ for
$x\in{\mathsf {\mathcal C}B}({\mathcal H})_+$,
and $\tau(x)=\infty$ for $x\in {\mathsf B}({\mathcal H})_+
\backslash {\mathsf {\mathcal C}B}({\mathcal H})_+$,
a non-zero trace $\tau$ on ${\mathsf B}({\mathcal H})$ is given.
Thus, traces (resp.~non-trivial traces)
on all bounded linear operators are in one-to-one correspondence
with traces (resp.~non-trivial traces) on the compact operators.

For the unique extension of the trace $\tr$ of Lemma \ref{spur.1} from
compact operators onto ${\mathsf B}({\mathcal H})$ the same notation
$\tr$ will be used. Note that in view of Lemma \ref{noncomp} with the help of
\ref{1.2.1} ($\times$) and ($\star\star$) easily follows that for non-compact
$x\geq {\mathbf 0}$ and each m.o.n.s. $\{\varphi_n\}$ one has
$\sum_{n=1}^\infty\langle x\varphi_n,\varphi_n\rangle=\infty$. Hence, the
formula given in Corollary \ref{spur} extends on all
$x\in {\mathsf B}({\mathcal H})_+$. From this formula
it is plain to see that $\tr$
is a non-trivial {\em normal} trace on ${\mathsf B}({\mathcal H})$.
Up to a positive multiple, $\tr$ is also unique on ${\mathsf B}({\mathcal H})$
as non-trivial trace with this property\,:
\begin{corolla}\label{unitra}
A non-trivial normal trace $\tau$ has
the form $\tau=\alpha\cdot\tr$, with $\alpha>0$.
\end{corolla}
\begin{proof}
Let $p_1<p_2<p_3<\ldots<{\mathbf 1}$ be a sequence of orthoprojections with
${\mathrm{rank}}(p_n)=n$, for each $n\in {\mathbb N}$. Then, for each
$x\geq {\mathbf 0}$, $l.u.b. \sqrt{x}\, p_n \sqrt{x}=x$.
Note that $x_n=\sqrt{x} p_n \sqrt{x}\in {\mathsf {\mathcal F}B}({\mathcal H})_+$
holds. Since also $\tau|_{{\mathsf {\mathcal C}B}({\mathcal H})_+}$ is a
non-trivial trace, by Corollary \ref{zerleg} there is unique $\alpha>0$ with
$\tau(x_n)=\alpha\cdot\tr x_n$, for each $n\in {\mathbb N}$.
Hence, by normality of $\tau$ and since $\tr$ is normal,
$\tau(x)=\alpha\cdot\tr x$ follows, for each $x\geq {\mathbf 0}$.
\end{proof}
Note that in view of the mentioned one-to-one correspondence with traces on the compact
operators Corollary \ref{zerleg} extends to non-trivial traces on
${\mathsf B}({\mathcal H})$ accordingly. In line with this and
Corollary \ref{unitra} the theory of traces on ${\mathsf B}({\mathcal H})$ with separable
infinite dimensional ${\mathcal H}$ essentially is the theory of the one
normal trace $\tr$ and myriads of singular traces.

\subsection{Examples of singular traces on ${\mathsf B}({\mathcal H})$}\label{1.3}
Examples of singular traces have been invented by J.\,Dixmier in
\cite{Dixm:66}. Nowadays this class is referred to as Dixmier-traces. In the following, only
the singular traces of this class will be constructed and considered. Thereby, in constructing
these traces we will proceed in two steps.

In a first
step we are going to define some non-trivial two-sided ideal in
${\mathsf B}({\mathcal H})$, with hereditary positive cone, which
later will prove to belong to the
defining ideal of each of the singular traces to be constructed. As has been already noticed
in context of Theorem \ref{calkin}, each such ideal then is an ideal of
compact operators. For such ideals one knows that these can be completely
described in terms of the classes (Schatten-classes) of the characteristic sequences
coming along with the operators of the ideal, see \cite[Theorem 12]{Scha:70}.
In these sequences, which are in $\ell^\infty(\mathbb{N})_+$,
the full information on the ideal is encoded.

In a second step, a class of
states on $\ell^\infty(\mathbb{N})$ is constructed which, in restriction
to the mentioned sequences from the ideal, yields a map which vanishes on
those sequences which correspond to operators of finite rank. If taken as functions
on the positive operators of the ideal these maps will be shown to be additive, positive
homogeneous and invariant. Hence, the extension
via the extension principle of Lemma \ref{nullaux} on all of
${\mathsf B}({\mathcal H})_+$ finally will provide us with a class of
singular traces.

\subsubsection{Step one\textup{:} Some ideal of compact operators}\label{1.3.1}
For compact $x$ with the help of the characteristic
sequence $\{\mu_n(x)\}$ define
\begin{subequations}\label{logdiv}
\begin{equation}\label{logdiv.1}
\forall\,k\in {\mathbb N}\backslash\{1\}:\ \gamma_k(x)=\frac{1}{\log k}\sum_{j\leq k} \mu_j(x)=\frac{\sigma_k(x)}{\log k}\,.
\end{equation}
Then, $\{\gamma_n(x):\,n>1\}$ is a sequence of non-negative reals which may be bounded or not.
The bounded situation deserves our special interest. Let a subset $L^{1,\infty}({\mathcal H})\subset
{\mathsf {\mathcal C}B}({\mathcal H})$  be defined as follows\,:
\begin{equation}\label{logdiv.2}
L^{1,\infty}({\mathcal H})=\left\{x\in {\mathsf {\mathcal C}B}({\mathcal H}):\ \sup_{n\geq 2}\gamma_n(x)<\infty\right\}\,.
\end{equation}
\end{subequations}
It is plain to see that by $L^{1,\infty}({\mathcal H})$ an ideal
is given in ${\mathsf B}({\mathcal H})$, for some corresponding terminology see
\cite{Maca:61,Conn:88,Conn:94}, and e.g.~\cite{GrVa:93}.
\begin{satz}\label{macaev}
$L^{1,\infty}({\mathcal H})$ is a non-trivial two-sided ideal in ${\mathsf B}({\mathcal H})$, and thus is
an ideal of compact operators, with hereditary cone $L^{1,\infty}({\mathcal H})_+$ of positive elements.
\end{satz}
\begin{proof}
In view of the definitions \eqref{logdiv}
and since ${\mathsf {\mathcal C}B}({\mathcal H})$ is a two-sided ideal,
the validity of the first assertion follows as an immediate consequence of \eqref{basis.1},
\eqref{basis.1b} and
\eqref{basis.2a} together with the fact that for each operator $x$ of finite rank
$\{\gamma_n(x):\,n>1\}$ is a
null-sequence and thus is bounded. Finally, owing to Corollary \ref{here} for
$x\in L^{1,\infty}({\mathcal H})_+$ and $y\in {\mathsf B}({\mathcal H})$ with ${\mathbf 0}\leq y\leq x$
one infers $y\in {\mathsf {\mathcal C}B}({\mathcal H})_+$, and then $y\leq x$ according to
\eqref{basis.1a} implies also $y\in L^{1,\infty}({\mathcal H})_+$.
\end{proof}

For completeness yet another characterization of
$L^{1,\infty}({\mathcal H})$ will be noted (without proof,
see e.g.~in \cite[IV.2.$\beta$]{Conn:94}), and a class of $L^{1,\infty}$-elements,
which can be characterized through the asymptotic behavior of the
singular values, will be given.

Let $L^1({\mathcal H})$ be the ideal of all operators of {\em trace-class}, that is, the defining ideal which
corresponds to the normal trace $\tr$, cf.~Lemma \ref{spur.1} and Corollary \ref{spur}. From \eqref{logdiv.2} and
Lemma \ref{spur.1} then especially follows that the inclusion relation
$L^1({\mathcal H})\subset L^{1,\infty}({\mathcal H})$ takes place amongst
$L^{1,\infty}({\mathcal H})$ and the ideal of trace-class operators.
Moreover, if in line with \cite{Maca:61} another Banach space
$L^{\infty,1}({\mathcal H})$ ($={\mathfrak S}_\omega$ in \cite{Maca:61}) is defined through
\begin{equation}\label{macaev.1}
L^{\infty,1}({\mathcal H})=\biggl\{y\in {\mathsf {\mathcal C}B}({\mathcal H}):\
\sum_{n=1}^\infty n^{-1}\mu_n(y)<\infty \biggr\}\,,
\end{equation}
then it is essentially due to \eqref{basis.1b}, \eqref{basis.2b} and by monotonicity of the sequences
of the $\sigma_n(y)$'s and $\frac{1}{n}$'s that also $L^{\infty,1}({\mathcal H})$
is a non-trivial two-sided ideal (Macaev-ideal).
Note that in analogy to the above also in this case obviously an inclusion with
trace-class operators takes place,
$L^1({\mathcal H})\subset L^{\infty,1}({\mathcal H})$. The ideals from
\eqref{logdiv.2} and \eqref{macaev.1} are related by the duality
given through the $2$-form $\varOmega(x,y)=\tr xy$. Namely, each
$x\in {\mathsf {\mathcal C}B}({\mathcal H})$ obeying $xy \in L^1({\mathcal H})$, for all
$y\in  L^{\infty,1}({\mathcal H})$, is in $ L^{1,\infty}({\mathcal H})$.
\begin{satz}\label{macaev.2}
$L^{1,\infty}({\mathcal H})$ is the dual to the Macaev-ideal.
\end{satz}
Also, in this context note that for
each $x\in {\mathsf {\mathcal C}B}({\mathcal H})$ obeying $xz \in L^1({\mathcal H})$ for all $z\in I$,
with an ideal $I$ of compact operators, and each $y\in I$ the relation
\begin{subequations}\label{macaev2}
\begin{equation}\label{macaev.2a}
|\varOmega(x,y)|\leq \sum_{n=1}^\infty \mu_n(x)\,\mu_n(y)<\infty
\end{equation}
must be fulfilled. In fact, since by assumption for $a,b\in {\mathsf B}({\mathcal H})$
also $axby\in L^1({\mathcal H})$ is fulfilled, in view of the polar decomposition
of $x,y$ the estimate
$|\varOmega(x,y)|\leq \sup_{u,v} |\tr u|x|v|y|\,|$ can be easily inferred, with $u,v$ extending over
the partial isometries in ${\mathsf B}({\mathcal H})$. Also,
with the help of Corollary \ref{spur}, and \eqref{sing.2} e.g., one finds
that $\sup_{u,v} |\tr u|x|v|y|\,|\leq
\sup_{\,\vec{r}}\sum_{n=1}^\infty \mu_n(x) r_n$ must hold, with $\vec{r}=(r_1,r_2,\ldots)$ obeying
$r_1\geq r_2\geq r_3\geq \ldots\geq 0$ and $\sum_{k\leq n} r_k\leq \sigma_n(y)=
\sum_{k\leq n} \mu_k(y)$,
for each $n\in {\mathbb N}$ (the ordering of $\mu_n(x)$'s is of importance in this context).
Since also the sequence of $\mu_n(y)$'s is in decreasing order, it is not hard to see that by
successively exploiting the just mentioned
conditions on $\vec{r}$, for $n\leq N$ with $N\in {\mathbb N}$, the validity of
$\sum_{k\leq N} \mu_k(x)(\mu_k(y)-r_k)\geq 0$
can be derived, for each $N\in {\mathbb N}$, and any given $\vec{r}$ which is
subject to the above conditions. From this the left-hand side estimate of \eqref{macaev.2a} gets evident.
Now, for any two given compact linear
operators $x,y$ in view of the polar decomposition theorem and owing to
compactness of both operators partial isometries $u,w$ can be chosen such that
$u|x|w|y|\geq {\mathbf 0}$ holds, with the singular values of the
compact operator $u|x|w|y|$ obeying $\mu_n(u|x|w|y|)=\mu_n(x)\,\mu_n(y)$,
for each $n\in {\mathbb N}$. In accordance with Lemma \ref{spur.1} one then has
$|\tr u|x|v|y|\,|=\tr u|x|w|y|=\lim_{n\to\infty} \sigma_n(u|x|w|y|)=\sum \mu_n(u|x|w|y|)=
\sum \mu_n(x)\,\mu_n(y)$. Hence, since in our particular situation of $x,y$ we have
$u|x|w|y|\in L^1({\mathcal H})$ and the above proved left-hand side estimate of \eqref{macaev.2a} has been
shown to hold, \eqref{macaev.2a} is completely seen.

Especially, in view of Proposition \ref{macaev.2} the estimate \eqref{macaev.2a} can be applied with
$x\in L^{1,\infty}({\mathcal H})$ and $I=L^{\infty,1}({\mathcal H})$. Relating asymptotic properties of
singular values of
$x\in L^{1,\infty}({\mathcal H})$ we thus get the following information\,:
\begin{equation}\label{macaev.4}
x\in  L^{1,\infty}({\mathcal H})\ \Longrightarrow\ \forall\,y\in L^{\infty,1}({\mathcal H})\,:\ \sum_{n=1}^\infty \mu_n(x)\,\mu_n(y)<\infty\,.
\end{equation}
\end{subequations}
Viewing \eqref{macaev.1} and \eqref{macaev.4} together suggests compact $x$ with asymptotic behavior
of singular values like $\mu_n(x)=
{\mathbf O}(n^{-1})$ as good candidates for elements of $L^{1,\infty}({\mathcal H})$.\footnote{As usual, for $g:{\mathbb N}\ni n\mapsto g(n)\in {\mathbb R}_+\backslash \{0\}$
the notation $x_n={\mathbf O}(g(n))$ is a shorthand notation for $|x_n|<C\,g(n)$, with some $C>0$ (and accordingly
defined with ${\mathbb R}_+$ instead
of ${\mathbb N}$).} In fact, such asymptotic behavior
implies that, with some $C>0$, for all $n\geq 2$
$$\sigma_n(x)=\sum_{1\leq k\leq n}\mu_k(x)\leq C\biggl\{1+\sum_{2\leq k\leq n} k^{-1}\biggr\}\leq C\biggl\{1+\int_1^n t^{-1} d\/t\biggr\}
=C(1+\log{n})$$
is fulfilled. In view of \eqref{logdiv} we therefore arrive at the following result\,:
\begin{corolla}\label{macaev.3}
$ x\in {\mathsf {\mathcal C}B}({\mathcal H}),\, \mu_n(x)=
{\mathbf O}(n^{-1})\  \Longrightarrow\ x\in L^{1,\infty}({\mathcal H})\,.$
\end{corolla}
\noindent
\begin{remark}\label{varilly0}
\begin{enumerate}
\item\label{varilly0.1}
It is easy to see that for compact $x$ with bounded multiplicity function,
$m(\lambda)\leq N<\infty$ for all $\lambda$, the condition imposed by \eqref{macaev.4} upon $x$
amounts to $\mu_n(x)={\mathbf O}(n^{-1})$.
Unfortunately, in case of unbounded $m$ this can fail to hold.\,\footnote{We are grateful to C.\,Portenier for
mentioning this fact to us.} That this can even occur for $x$ within
$L^{1,\infty}({\mathcal H})$ can be seen by the following counterexample\,:\,\footnote{The counterexample has been communicated
to us by J.~V\'{a}rilly,
see also \cite[Lemma 7.35]{Vari:2000}.}
\item\label{varilly0.2}
Let $x$ be positive and compact with $\mu_1(x)=1$, and with singular values which for $k\geq 2$ with
$(m-1)!<k\leq m!$, $m\geq 2$, are given by $\mu_k(x)=\log{m}/\,m!$.
One then easily proves that $\sigma_k(x)$ obeys $\sigma_{m!}(x)\leq 1+\log{m!}$.
Since the function $f(t)=\log(1+t/m!)-t\{\log(m+1)/\,(m+1)!\}$ is non-negative for $0\leq t\leq m\cdot m!$,
from the
previous also $\sigma_k(x)\leq 1+\log{k}$ can be followed
whenever $m!<k\leq (m+1)!$ is fulfilled. This conclusion applies
for each $m\geq 2$, and thus according to \eqref{logdiv} we finally get
$x\in  L^{1,\infty}({\mathcal H})$. On the other hand, $\lim_{m\to\infty} m!\,\mu_{m!}(x)=\infty$ holds.
Thus in particular $\mu_n(x)$ certainly cannot behave asymptotically like
${\mathbf O}(n^{-1})$.
\end{enumerate}
\end{remark}

\subsubsection{Step two\textup{:} Scaling invariant states}\label{1.3.2}
Let us come back now to the construction of the Dixmier-traces. The construction will be based on considering a
certain class of states on the
commutative
${\mathsf W^*}$-algebra $M=\ell^\infty$, see \ref{1.1.1} for basic notations.
Relating special further notations, for each $k\in {\mathbb N}$ let $e_k\in \ell^\infty$
be the $k$-th atom in
$\ell^\infty$, with $j$-th component obeying $(e_k)_j=\delta_{kj}$ (Kronecker
symbol), and let $E_k$ be the special orthoprojection of rank $k$ given as $E_k=\sum_{j\leq k} e_j$.
The ascendingly directed sequence $\{E_n\}$ obeys $\mathop{l.u.b.} E_n={\mathbf 1}$ and
the following equivalence is valid\,:
\begin{equation}\label{aux.1}
x\in \ell^\infty:\ \|\cdot\|_\infty-\lim_{n\to\infty} E_n x =x\ \Longleftrightarrow\ \lim_n x_n=0\,.
\end{equation}
Also, for $x\in \ell_+^\infty$, $\{E_n x\}\subset\ell_+^\infty $ is ascendingly
directed, with $\mathop{l.u.b.} E_n x=x$.

For the following, let a mapping ${\mathbf s}:\ell^\infty\longrightarrow \ell^\infty$ (scaling)
be defined on
$x\in \ell^\infty$ through ${\mathbf s}(x)_j=x_{2j}$, for all $j\in {\mathbb N}$.
It is obvious that ${\mathbf s}$ is a normal $^*$-homomorphism onto $\ell^\infty$.
Hence, ${\mathbf s}$ is a unital normal positive linear map onto itself, and
$\ell_{\mathbf s}^\infty=\{x\in \ell^\infty:{\mathbf s}(x)=x\}$ is a
${\mathsf W}^*$-subalgebra of $\ell^\infty$ (the
fixpoint algebra of ${\mathbf s}$).
\begin{lemma}\label{scaling}
There exists a conditional expectation
${\mathcal E}:\ell^\infty\,\longrightarrow\,\ell_{\mathbf s}^\infty$ projecting onto
the fixpoint algebra $\ell_{\mathbf s}^\infty$ such that the following
properties hold\textup{:}
\begin{enumerate}
\item\label{scaling.1}
${\mathcal E}\circ {\mathbf s}={\mathcal E}\,;$
\item\label{scaling.3}
${\mathcal E}(x)=(\lim_{n\to\infty} x_n)\cdot {\mathbf 1}$, for each $x\in\ell^\infty$ with
$\lim_{n\to\infty} x_n$ existing.
\end{enumerate}
\end{lemma}
\begin{proof}
Let us consider the sequence
$\{{\mathbf s}\langle n\rangle\}$ of partial averages ${\mathbf s}\langle n\rangle=\frac{1}{n}\sum_{k\leq n}
{\mathbf s}^k$, $n\in {\mathbb N}$. Since these all are unital positive linear
maps, by $\sigma(\ell^\infty,\ell^1)$-weak compactness of the closed
unit ball in ${\mathsf B}(\ell^\infty)$
the sequence of partial averages then must have a $\sigma(\ell^\infty,\ell^1)$-weak cluster
point ${\mathcal E}$ which has to be a unital positive linear map, too.
Since then ${\mathcal E}=\sigma(\ell^\infty,\ell^1)-\text{weak } \lim_\lambda
{\mathbf s}\langle n_\lambda\rangle$
has to be fulfilled for some appropriately chosen
subnet $\{{\mathbf s}\langle n_\lambda\rangle\}$, the inclusion
$\ell_{\mathbf s}^\infty\subset\{x\in \ell^\infty:{\mathcal E}(x)=x\}$ gets
evident.
Since ${\mathbf s}\langle n\rangle\circ {\mathbf s}=
{\mathbf s}\circ {\mathbf s}\langle n\rangle$ and
$\|{\mathbf s}\langle n\rangle\circ {\mathbf s}-{\mathbf s}\langle n\rangle\|\leq \frac{2}{n}$ hold,
for each $n\in {\mathbb N}$, and since owing to normality of ${\mathbf s}$
for each $\omega\in \ell^1$ also $\omega\circ {\mathbf s}\in \ell^1$
is fulfilled, by argueing with the mentioned subnet
one infers that ${\mathcal E}\circ {\mathbf s}={\mathbf s}\circ {\mathcal E}={\mathcal E}$. From this
$\{x\in \ell^\infty:{\mathcal E}(x)=x\}\subset\ell_{\mathbf s}^\infty$ and
${\mathbf s}\langle n\rangle\circ {\mathcal E}={\mathcal E}$ follow, for each $n$. Thus
in view of the above ${\mathcal E}^2={\mathcal E}\circ {\mathcal E}={\mathcal E}$ follows.
Hence, ${\mathcal E}$ is a projection of norm one
(conditional expectation) projecting onto the
fixpoint algebra of ${\mathbf s}$ and which satisfies (\ref{scaling.1}).

To see (\ref{scaling.3}), note first that owing to ${\mathbf s}(e_k)={\mathbf 0}$ for $k$ odd, and
${\mathbf s}(e_k)=e_{k/2}$ for $k$ even, one certainly has ${\mathbf s}^n(E_k)={\mathbf 0}$,
for each $n>\log k / \log 2$. Hence, the action of the $n$-th average ${\mathbf s}\langle n\rangle$ to the orthoprojection
$E_k$ can be estimated as $\|{\mathbf s}\langle n\rangle(E_k)\|_\infty
\leq [\log k / \log 2]/n$ (here $[\cdot]$ means the integer part), and thus for all
$k\in {\mathbb N}$ one has
$\|\cdot\|_\infty-\lim_{n\to\infty} {\mathbf s}\langle n\rangle(E_k)=
{\mathbf 0}$.
>From this and ${\mathcal E}=\sigma(\ell^\infty,\ell^1)-\text{weak } \lim_\lambda
{\mathbf s}\langle n_\lambda\rangle$ then especially $\omega({\mathcal E}(E_k))=0$ follows, for each $\omega\in \ell^1$. Hence
${\mathcal E}(E_k)={\mathbf 0}$, for each $k$.
Since for each $y\in \ell^\infty$ with ${\mathbf 0}\leq y\leq {\mathbf 1}$
one has ${\mathbf 0}\leq E_k y\leq E_k$, from the previous together with
positivity of ${\mathcal E}$ also
${\mathcal E}(E_k y)={\mathbf 0}$ follows. By linearity of ${\mathcal E}$ and
since $\ell^\infty$ is the linear span of
$\ell_+^\infty \cap(\ell^\infty)_1$ this remains true for each
$y\in \ell^\infty$. But then, for $x\in \ell^\infty$ with
$\alpha=\lim_{n\to\infty} x_n$ by continuity of ${\mathcal E}$ and in view of
(\ref{aux.1}) one infers
${\mathcal E}(x-\alpha\cdot{\mathbf 1})=
\|\cdot\|_\infty-\lim_{k\to\infty} {\mathcal E}(E_k(x-\alpha\cdot{\mathbf 1}))=
{\mathbf 0}$, which is equivalent with (\ref{scaling.3}).
\end{proof}
\begin{corolla}\label{dix}
There is a state $\omega\in {\mathcal S}(\ell^\infty)$ satisfying
the following properties\textup{:}
\begin{enumerate}
\item\label{dix.1}
$\omega\circ {\mathbf s}=\omega\,;$
\item\label{dix.3}
$\omega(x)=\lim_{n\to\infty} x_n$, provided
$\lim_{n\to\infty} x_n$ exists.
\end{enumerate}
The set $\Gamma_{\mathsf s}(\ell^\infty)$ of all such states is a $w^*$-compact convex subset of
singular states.\,\footnote{Let another map
${\mathbf d}$ (doubling) over $\ell^\infty$ be defined at $x$ by ${\mathbf d}(x)_j=x_{[(1+j)/2]}$,
$j\in {\mathbb N}$ ($[r]$ refers to the integer part of $r$). Then,
scaling is left-inverse to doubling in ${\mathsf B}(\ell^\infty)$,
and thus in addition to \eqref{dix.1} one also has ${\mathbf d}$-invariance of each
$\omega\in\Gamma_{\mathsf s}(\ell^\infty)$ as well.}
\end{corolla}
\begin{proof}
Let ${\mathcal E}$ be constructed as in Lemma \ref{scaling}. By positivity and unitality of ${\mathcal E}$, for each
$\nu\in {\mathcal S}(\ell^\infty)$ also
$\omega=\nu\circ {\mathcal E}$ is a state. In view of (\ref{scaling.1})--(\ref{scaling.3}) this state
then obviously satisfies (\ref{dix.1})--(\ref{dix.3}).\footnote{The usage of Lemma \ref{scaling}
might be avoided in this context; as we learned from \cite{schm}
a positivity and separation argument of Hahn-Banach type may be used instead as well.}
That $\Gamma_{\mathsf s}(\ell^\infty)$ is $w^*$-compact and
convex is evident from the linear nature of the conditions (\ref{dix.1})--(\ref{dix.3}). Finally,
in accordance with (\ref{dix.3}) one has $\omega(E_k)=0$, for each
$\omega\in \Gamma_{\mathsf s}(\ell^\infty)$ and all $k\in {\mathbb N}$.
Now, let $p\in \ell^\infty$ be any orthoprojection
with $\omega(p)>0$. Then, $p\not={\mathbf 0}$, and owing to
${\mathop{l.u.b.} E_n p}=p$ there has to exist
$k\in {\mathbb N}$ with $q=E_k p\not={\mathbf 0}$. Thus ${\mathbf 0}<q<p$
and $q\leq E_k$. In view of the above from the latter
by positivity of $\omega$ then $\omega(q)=0$ follows. Hence, each
$\omega\in \Gamma_{\mathsf s}(\ell^\infty)$ is singular.
\end{proof}

\subsubsection{Constructing the Dixmier-traces}\label{1.3.3}
For given $x\in L^{1,\infty}({\mathcal H})$, let a sequence
$\gamma(x)$ be given through
$\gamma(x)=(\gamma_2(x),\gamma_3(x),\ldots )$, with $\gamma_n(x)$ in accordance
with \eqref{logdiv.1}. Then, by definition \eqref{logdiv.2} one has
$\gamma(x)\in\ell_+^\infty$. Hence, if for each fixed scaling invariant state
$\omega\in \Gamma_{\mathsf s}(\ell^\infty)$, see Corollary \ref{dix}, following \cite{Dixm:66} we define
\begin{equation}\label{dixmier}
\forall\,x\in L^{1,\infty}({\mathcal H})_+\,:\ {\Tr}_\omega(x)=\omega(\gamma(x))\,,
\end{equation}
then according to Proposition \ref{macaev} and since $\omega$ is a
positive linear form, we are given a {\em positive} map
${\Tr}_\omega:L^{1,\infty}({\mathcal H})_+\ni x\,\longmapsto {\Tr}_\omega(x) \in
{\mathbb R}_+$ defined on the positive cone of the ideal $L^{1,\infty}({\mathcal H})$.
The key idea of \cite{Dixm:66} is that additivity of ${\Tr}_\omega$ can be shown.
\begin{lemma}\label{lini}
${\Tr}_\omega$ is an additive, positive homogeneous and invariant map from
$L^{1,\infty}({\mathcal H})_+$ into ${\mathbb R}_+$.
\end{lemma}
\begin{proof}
Since $L^{1,\infty}({\mathcal H})_+$ is the positive cone of a two-sided ideal
of compact operators, for $x,y\in L^{1,\infty}({\mathcal H})_+$ and
$\lambda\in {\mathbb R}_+$ we have that
$x+y,\,\lambda\,x,\,x^*x,\,xx^*\in L^{1,\infty}({\mathcal H})_+$, and these are
compact operators again. Hence, in view of \eqref{logdiv.1} from
\eqref{basis.1} and \eqref{basis.3} both
$\lambda\cdot\gamma(x)=\gamma(\lambda\,x)$ and
$\gamma(x^*x)=\gamma(xx^*)$ follow, which in line with \eqref{dixmier} means
that ${\Tr}_\omega$ is positive homogeneous and invariant.
It remains to be shown that ${\Tr}_\omega$ is additive.
First note that according to the
left-hand side estimate of \eqref{basis.2} within $\ell_+^\infty$
one has $\gamma(x+y)\leq \gamma(x)+\gamma(y)$. Hence, by positivity and linearity of $\omega$,
\eqref{dixmier} yields
\begin{equation}\tag{$\star$}
{\Tr}_\omega (x+y)\leq {\Tr}_\omega(x) +{\Tr}_\omega(y)\,.
\end{equation}
Now, to each compact operator $z$ let $\gamma^0(z)=(\gamma_3(z),\gamma_4(z),\ldots)$, that is,
$\gamma^0(z)$ arises from $\gamma(x)$ by application of the one-step left-shift. Also,
on $\ell^\infty$ let a linear map ${\mathbf m}$ be defined by ${\mathbf m}(\beta)_n=
\frac{\log 2}{\log (n+1)}\cdot\beta_n$, for all $n\in {\mathbb N}$, at $\beta\in \ell^\infty$. One
then has
\begin{equation}\tag{$\star\star$}
\forall\,\beta\in \ell^\infty\,:\ \lim_{n\to\infty}{\mathbf m}(\beta)_n=0\,.
\end{equation}
Note that
$\gamma^0(z)\in \ell^\infty$ whenever $z\in L^{1,\infty}({\mathcal H})$. We are going to
estimate $\gamma(z)-\gamma^0(z)$ for $z\in L^{1,\infty}({\mathcal H})$. Since both $\{\sigma_n(z)\}$
and $\{\log n \}$ are monotoneously increasing, in view of the definition \eqref{logdiv.1} for
each $z\in L^{1,\infty}({\mathcal H})$
the follwing estimates at once can be seen to hold, for all $k\in
{\mathbb N}\backslash \{1\}$\,:
$$
\gamma_k(z)-\gamma_{k+1}(z)\leq \gamma_{k+1}(z)\biggl(\frac{\log(k+1)}{\log k}-1\biggr)\leq
\frac{\log 2}{\log k}\cdot\gamma_{k+1}(z)\leq \frac{\log 2}{\log k}\,\|\gamma(z)\|_\infty\,.
$$
On the other hand, we also have
$$
\gamma_k(z)-\gamma_{k+1}(z)\geq \frac{\sigma_k(z)-\sigma_{k+1}(z)}{\log (k+1)}=
-\frac{\mu_{k+1}(z)}{\log (k+1)}\geq -\frac{1}{\log (k+1)}\cdot \|z\|\,.
$$
>From these two estimates we infer that $\Delta(z)=\gamma(z)-\gamma^0(z)$ for
$z\in L^{1,\infty}({\mathcal H})$ is a null-sequence in $\ell^\infty$, that is,
$\lim_{n\to\infty} \Delta(z)_n=0$ is fulfilled. According to the choice of $\omega$ and in
accordance with Corollary \ref{dix}\,\eqref{dix.3} we thus have the following to hold\,:
\begin{equation}\tag{$\star\star\star$}
\forall\,z\in L^{1,\infty}({\mathcal H}),\,\omega\in \Gamma_{\mathsf s}(\ell^\infty):\
\omega(\gamma(z))=\omega(\gamma^0(z))\,.
\end{equation}
Let us come back to our above $x,y\in L^{1,\infty}({\mathcal H})_+$.
Having in mind the definitions of the positive linear operators
${\mathbf s}$ and ${\mathbf m}$ as well as the meanings of
$\gamma$ and $\gamma^0$, it is easily inferred
that from the
right-hand side estimate in \eqref{basis.2} when divided by $\log k$,
and considered for all $k\geq 2$,
the estimate
$\gamma(x)+\gamma(y)\leq (\sigma_{2k}(x+y)/\log k)=
{\mathbf s}(\gamma^0(x+y))+{\mathbf m}\circ{\mathbf s}(\gamma^0(x+y))$
can be followed to hold in
$\ell_+^\infty$.
By positivity and linearity of $\omega$
from this then
\begin{equation}\tag{$\circ$}
\omega(\gamma(x))+\omega(\gamma(y))\leq
\omega\circ{\mathbf s}(\gamma^0(x+y))+\omega\circ{\mathbf m}({\mathbf s}(\gamma^0(x+y)))
\end{equation}
follows.
Now, in view of ($\star\star$) and Corollary \ref{dix}\,\eqref{dix.3} one has
$\omega\circ{\mathbf m}({\mathbf s}(\gamma^0(x+y)))=0$, whereas from
Corollary \ref{dix}\,\eqref{dix.1}
and ($\star\star\star$) one concludes that
$\omega\circ{\mathbf s}(\gamma^0(x+y))=\omega(\gamma(x+y))$.
These facts together with ($\circ$) fit together into the estimate
$\omega(\gamma(x))+\omega(\gamma(y))\leq
\omega(\gamma(x+y))$, which in view of \eqref{dixmier} says that ${\Tr}_\omega(x) +{\Tr}_\omega(y)
\leq {\Tr}_\omega (x+y)$ has to be valid. The latter and ($\star$) then make that
the desired additivity ${\Tr}_\omega(x) +{\Tr}_\omega(y)
={\Tr}_\omega (x+y)$ holds.
\end{proof}
\subsubsection{The Dixmier-trace as a singular trace}\label{1.3.4}
We remind that according to Proposition \ref{macaev} the positive cone of the ideal
$L^{1,\infty}({\mathcal H})$ is hereditary. Thus the
extension principle of Lemma \ref{nullaux} according to
Lemma \ref{lini} for each $\omega\in \Gamma_{\mathsf s}(\ell^\infty)$ allows to extend the
map ${\Tr}_\omega$ of \eqref{dixmier} to a trace on
${\mathsf B}({\mathcal H})$. Thereby, the extension constructed in accordance with the proof of Lemma \ref{nullaux}
will be the unique one with defining ideal $L^{1,\infty}({\mathcal H})$.
For this trace the same notation ${\Tr}_\omega$ will be used henceforth. We refer to this trace
as {\em Dixmier-trace} (to the particular $\omega\in \Gamma_{\mathsf s}(\ell^\infty)$).
The essential properties of Dixmier-traces are summarized in the following.
\begin{theorem}\label{ex}
${\Tr}_\omega$ is a singular
trace on ${\mathsf B}({\mathcal H})$, for each $\omega\in \Gamma_{\mathsf s}(\ell^\infty)$.
The following properties are fulfilled\textup{:}
\begin{enumerate}
\item\label{ex.1}
$L^{1,\infty}({\mathcal H})_+=\{x\in {\mathsf B}({\mathcal H})_+:\ {\Tr}_\omega(x)<\infty\}$\,;
\item\label{ex.2}
$x\in L^{1,\infty}({\mathcal H})_+,\,\exists\,\lim_{n\to\infty}\gamma_n(x)\ \Longrightarrow\
{\Tr}_\omega(x)=\lim_{n\to\infty}\gamma_n(x)$\,.
\end{enumerate}
\end{theorem}
\begin{proof}
The validity of \eqref{ex.1} follows since the traces in question all
are obtained
as extensions of the maps given in \eqref{dixmier}, which satisfy Lemma \ref{lini} and which have range
${\mathbb R}_+$ (and not merely $\overline{{\mathbb R}}_+$). Since each state
$\omega\in \Gamma_{\mathsf s}(\ell^\infty)$ obeys Corollary \ref{dix}\,\eqref{dix.3}, in view of the previous and
\eqref{dixmier} also \eqref{ex.2} follows. Finally, for each
$x\in {\mathsf {\mathcal F}B}({\mathcal H})_+$ the sequence $\gamma(x)$ is a null-sequence, and
therefore especially $x\in L^{1,\infty}({\mathcal H})_+$, and as a special case of
\eqref{ex.2} then ${\Tr}_\omega(x)=0$ follows. Hence, ${\Tr}_\omega$ is a singular trace.
\end{proof}
Note the remarkable feature of the Dixmier-traces coming along with Theorem \ref{ex}\,\eqref{ex.2}
and saying that provided certain circumstances are fulfilled for $x$, e.g.~if the sequence
$\{\gamma_n(x)\}$ has a limit, then independent of the state-parameter
$\omega$ all these Dixmier-traces may yield the same common value at this $x$. It is such case of independence one
usually is tacitely addressing to
when speaking simply of {\em the} Dixmier-trace of $x$, whereas the operator itself then is
referred to as {\em measurable operator}, cf.~\cite[IV.2,\,Definition 7]{Conn:94}. Some criteria of
measurability, which however all reduce upon showing that the above mentioned special case of
existence of $\lim_{n\to\infty}\gamma_n(x)$ would happen, subsequently will be discussed in more detail.

\subsection{Calculating the Dixmier-trace}\label{1.4}
\subsubsection{Simple criteria of measurability}\label{1.4.1}
We start with discussing conditions which
read in terms of spectral theory and which ensure that\,--\,for a given
operator $x\in L^{1,\infty}({\mathcal H})$ which is not simply of finite rank\,--\,the
above-mentioned special case of measurability occurs, that is, the
limit $\lim_{n\to\infty}\gamma_n(x)$ exists. As a first result of that kind one has the
following one:\footnote{We are grateful to C.\,Portenier,
Marburg, for suggesting some details around this and related subjects \cite{Port:99}.}
\begin{lemma}\label{ike4}
Suppose $x\in {\mathcal C}{\mathsf B}({\mathcal H})$, with $\mu_n(x)\sim L\cdot n^{-1}$. Then
$\lim_{n\to\infty}\gamma_n(x)=L$.\footnote{For $f:{\mathbb N}\rightarrow {\mathbb R}_+$ and
$g:{\mathbb N}\rightarrow {\mathbb R}_+\backslash \{0\}$
the notation $f(n)\sim L\cdot g(n)$ stands for $\lim_{n\to\infty} f(n)/g(n)=L$ (and accordingly
defined with ${\mathbb R}_+$ instead
of ${\mathbb N}$).}
\end{lemma}
\begin{proof}
For compact operator $x$ suppose $\lim_{n\to\infty} n\,\mu_n(x)=L$ to be
fulfilled. Then, in case of $L>0$,
for $\delta$ with $L>\delta>0$, let $M(\delta)\in {\mathbb N}$ be chosen
such that
\begin{equation*}
\forall\,n\geq M=M(\delta):\ (L-\delta)\,n^{-1}\leq \mu_n(x)\leq (L+\delta)\,n^{-1}\,.
\end{equation*}
>From this for each $n\geq M$ we get
\begin{equation*}\tag{$\times$}
(L-\delta)\sum_{M<k\leq n} k^{-1}\leq \sum_{M<k\leq n}\mu_n(x)\leq (L+\delta)\sum_{M< k\leq n} k^{-1}\,.
\end{equation*}
Since $0<t\,\mapsto\,t^{-1}$ is a strictly monotone decreasing function and
the sequence of the singular values is decreasingly ordered, with the help
of
$$
\int_{M+1}^n d\/t\,t^{-1} \leq\sum_{M<k\leq n} k^{-1}\leq \int_M^n d\/t\,t^{-1}
$$
which holds for $n>M$ the above estimate ($\times$) implies
\begin{equation*}
\int_{M+1}^{n} d\/t\,(L-\delta)\,t^{-1} \leq \sigma_n(x)-\sigma_M(x)\leq \int_M^{n} d\/t\,(L+\delta)\,t^{-1}\,.
\end{equation*}
>From this for all $n> M=M(\delta)$
$$(L-\delta)\{1-\log(M+1)/\log{n}\}\leq \gamma_n(x)-\sigma_M(x)/\log{n}\leq
(L+\delta)\{1-\log{M}/\log{n}\}$$
is obtained. Considering these estimates for $n\to\infty$ then yields
$$(L-\delta)\leq \liminf_{n\to\infty}\gamma_n(x)\leq \limsup_{n\to\infty}\gamma_n(x)\leq L+\delta\,.$$
Note that in case of $L=0$ by positivity of all $\gamma_n(x)$ instead of the previous one finds
$0\leq \liminf_{n\to\infty}\gamma_n(x)\leq \limsup_{n\to\infty}\gamma_n(x)\leq \delta$, for any $\delta>0$.
Thus, since $\delta>0$ can be chosen arbitrarily small,
in either case $\lim_{n\to\infty}\gamma_n(x)=L$ follows.
\end{proof}
Now, let us suppose $x\in {\mathcal C}{\mathsf B}({\mathcal H})$, with $\mu_n(x)={\mathbf O}(\frac{1}{n})$.
According to Corollary \ref{macaev.3} we even have $x\in L^{1,\infty}({\mathcal H})$, and
since $\sum_n  n^{-(1+\varepsilon)}<\infty$ is fulfilled
for each $\varepsilon>0$, then $|x|^z$ at each $z\in {\mathbb C}$ with $\Re z >1$ has to be of trace-class
and the definition
\begin{equation}\label{wiener.2}
\zeta_x(z)=\sum_{n\geq 1} \mu_n(x)^z=\tr |x|^z
\end{equation}
will provide us with some holomorphic function $\zeta_x$ in the half-plane $\Re z>1$. For this modification
of the Riemann $\zeta$-function the following holds.
\begin{lemma}\label{ike0}
Let $x\in {\mathcal C}{\mathsf B}({\mathcal H})$, with $\mu_n(x)={\mathbf O}(\frac{1}{n})$. Suppose $\zeta_x$ admits
an extension onto the half-plane
$\Re z\geq 1$ which is continuous there except for a simple pole with residue $L$ at $z=1$, at worst.
Then even $\mu_n(x)\sim L\cdot n^{-1}$ holds.
\end{lemma}
\begin{proof}
In case of $x\in {\mathsf {\mathcal F}B}({\mathcal H})$ one has $\lim_n n\cdot\mu_n(x)=0$ as well as
$\lim_{\varepsilon\to 0+} \tr |x|^{1+\varepsilon}=\tr |x|=\sum_n \mu_n(x)< \infty$, by triviality. From the
latter $\lim_{s\to 1+} (s-1)\,\zeta_x(s)=0$ follows. Hence, for each operator $x$ of finite rank the
assertion is true, with $L=0$.

Suppose now that $x$ is not of finite rank, $x\not\in {\mathsf {\mathcal F}B}({\mathcal H})$.
In view of definitions \eqref{logdiv.1} and \eqref{wiener.2}, upon possibly considering instead of $x$
a scaling $\lambda x$ by a suitably chosen real $\lambda>0$, without loss of generality
it suffices if the assertion for $x\geq {\mathbf 0}$ with $\mu_1(x)<1$ can be shown. In line with this
assume such  $x\in {\mathcal C}{\mathsf B}({\mathcal H})_+\backslash\, {\mathsf {\mathcal F}B}({\mathcal H})$.

By the spectral theorem there exists a spectral representation of $x$ as an operator Stieltjes-integral
$x=\int_0^{1-} \lambda\,E(d\/\lambda)$,
with projection-valued measure $E(d\/\lambda)$ derived from a left-continuous spectral family
$\{E(\lambda):\lambda\in {\mathbb R}\}$, that is, a family of orthoprojections obeying
$E(t)\leq E(\lambda)$, for $t\leq \lambda$, $E(s)={\mathbf 0}$, for $s\leq 0$ and
$E(\lambda-)=l.u.b._{t<\lambda} E(t)=E(\lambda)$, for each $\lambda\leq\infty$, with
$l.u.b._{t<\infty} E(t)={\mathbf 1}$. By convention, for $a<b$, then
$\int_a^{b-} E(d\/\lambda)=E(b)-E(a)=E([a,b[\,)$ and
$\int_a^b E(d\/\lambda)=E(b+)-E(a)=E([a,b])$, and so on accordingly,
where e.g.~$E(b+)$ stands for the greatest lower bound
$E(b+)=g.l.b._{t>b} E(t)$.

By means of some functional calculus and owing to
normality of the trace $\tr$ it is easily inferred that \eqref{wiener.2} can be represented as an ordinary
Stieltjes integral\,:
\begin{subequations}\label{ike}
\begin{equation}\label{ike.1}
\forall\,z\in {\mathbb C}, \Re z>1\,:\ \zeta_x(z)=\int_{1+}^\infty t^{-z}\,d\/\alpha(t)\,,
\end{equation}
with the monotone increasing function $\alpha$ given by
\begin{equation}\label{ike.2}
\alpha(t)=\tr{E([1/t,\infty[)}\,.
\end{equation}
But then, if the assumptions on $\zeta_x$ are fulfilled with
$\lim_{s\to 1+} (s-1)\,\zeta_x(s)=L$, all conditions for an application of Ikehara's theorem
\cite{Ikeh:31}, are given (we refer to the formulation in \cite[{\sc Theorem 16}]{Wien:60}).
In line with this the conclusion is that asymptotically
\begin{equation}\label{ike.3}
\alpha(t)\sim L\cdot t
\end{equation}
has to be fulfilled as $t$ tends to infinity.
\end{subequations}
Since $x$ is a compact operator, in view of the properties of the spectral resolution $E$
together with normality of $\tr$ the definition \eqref{ike.2} provides a right-continuous, integral-valued
step function which is constant between inverses of neighbouring
spectral values of $x$. Especially, in case of $n\in {\mathbb N}$ with
$\mu_n(x)>\mu_{n+1}(x)$  one infers that $\alpha$ for all $t$ with $\mu_n(x)^{-1}\leq t<\mu_{n+1}(x)^{-1}$ yields
$\alpha(t)=n$. A moments reflection then shows that with respect to each term of
the ordered sequence $n_1<n_2<n_3<\ldots $
of all subscripts where the value
of $\mu_n$ jumps the relation \eqref{ike.3} in view of $\lim_{n\to\infty} \mu_n(x)=0$ and by continuity of
the parameter $t$ in particular also implies
both $\lim_{k\to\infty} n_{k+1}\,\mu_{n_{k+1}}(x)=L$ and $\lim_{k\to\infty} n_k\,\mu_{n_{k+1}}(x)=L$ to be fulfilled.
But then, since $\mu_n(x)=\mu_{n_{k+1}}(x)$ holds for $n_k< n\leq n_{k+1}$, also
$\lim_{n\to\infty} n\,\mu_n(x)=L$ can be obtained from these limit relations.
Thus under the condition of the hypothesis
also $\mu_n(x)\sim L\cdot n^{-1}$ in case of $x\geq {\mathbf 0}$ and which is not of finite rank. In accordance
with our preliminary remarks the assertion then has to be true,
in either case under the mentioned hypothesis.
\end{proof}
\begin{remark}\label{resrem}
\begin{enumerate}
\item\label{resrem.1}
Relating Lemma \ref{ike4} remark that there are examples of
operators where $\lim_{n\to\infty} \gamma_n(x)=L$ exists but $\mu_n(x)\not\sim L\cdot n^{-1}$, see
\cite[Beispiel A.27]{Warz:97} or \cite[Lemma 7.35]{Vari:2000}.
\item\label{resrem.2}
On the one hand, the conditions imposed on $\mu_n(x)$ and $\zeta_x$ in Lemma \ref{ike0}
simply reproduce the usual conditions for the standard results of
Tauberian type\footnote{This especially concerns theorems of Hardy and Littlewood \cite{HaLi:18}
and Ikehara \cite{Ikeh:31}, see
\cite[Chap.\,VII,\,7.5]{Hard:49} and
\cite[see especially on p.\,126 and {\sc Theorem 18}]{Wien:60}.} to become applicable.
On the other hand, that the behavior of the extension of $\zeta_x$ at the whole line $\Re z=1$ (and not only at $z=1$)
has to be of relevance can be seen also by example\,: there is $x\in {\mathcal C}{\mathsf B}({\mathcal H})$ with $\mu_n(x)={\mathbf O}(\frac{1}{n})$ and $\lim_{s\to 1+} (s-1)\zeta_x(s)=1$ but for which
$\mu_n(x)\not\sim 1/n$.\footnote{J.~V\'{a}rilly has informed us about this fact and examples and counterexamples
around this question which will appear in \cite{Vari:2000}. Also we are very indebted to J.~V\'{a}rilly for some
clarifying remarks and hints to the literature.}
\end{enumerate}
\end{remark}
\subsubsection{A residue-formula for the Dixmier-trace}\label{1.4.2}
The most important from practical point of view special
case of measurability for an operator $x$ occurs if the limit $\lim_n \gamma_n(x)$ exists.
In particular, according to the
previous considerations the latter will happen e.g.~provided some function-theoretic assumptions on $x$
can be satisfied. In these
cases a formula arises which allows us to calculate
the (singular) Dixmier-trace with the help of the ordinary
trace as a limit of some function-theoretic expression of the operator in question.
In fact, in view of Theorem \ref{ex}\,\eqref{ex.2} and upon combining
Lemma \ref{ike0} and Lemma \ref{ike4} we get the following result\,:
\begin{corolla}\label{connes1}
For each $x\in {\mathcal C}{\mathsf B}({\mathcal H})_+$ with $\mu_n(x)={\mathbf O}(\frac{1}{n})$ one has
$x\in L^{1,\infty}({\mathcal H})_+$, and then by $\zeta_x(z)=\tr x^z$ a holomorphic function in the half-plane
$\Re z> 1$ is given. Suppose $\zeta_x$ extends onto the half-plane
$\Re z \geq 1$ and is continuous there except for a simple pole at $z=1$, at worst. Then
the Dixmier-trace of $x$ is obtained as
\begin{subequations}\label{connprop4}
\begin{equation}\label{prop41}
{\Tr}_\omega(x)=\lim_n \gamma_n(x)=\lim_{s\to 1+} (s-1)\tr x^s\,.
\end{equation}
Especially, when $\zeta_x$ extends to a meromorphic function on the whole complex plane, with
a simple pole at $z=1$ at worst, this formula turns into
\begin{equation}\label{propconn}
{\Tr}_\omega(x)={\mathop{\mathfrak{Res}}}|_{z=1}(\zeta_x)\,,
\end{equation}
\end{subequations}
with the residue ${\mathop{\mathfrak{Res}}}(\zeta_x)$
of the extended complex function, taken at $z=1$.
\end{corolla}
For completeness remark that by our Corollary \ref{connes1}, which is
sufficient to cope with our later needs around Connes' trace theorem, in the special cases at hand
the implication $(1)\Rightarrow (2)$ of \cite[IV,\,Proposition 4]{Conn:94} is reproduced.

Clearly, from both the theoretical and practical point of view, in context of the previous those
situations deserve the main interest where formula \eqref{propconn} could be applied. According to the results
in \cite[{\sc Theorem 7.1,\,7.2}]{Guil:85} this
happens e.g.~if the
context of the classical pseudodifferential operators of order $-n$ acting on the sections $\Gamma(E)$
of a complex vector bundle $E\rightarrow M$ of a $n$-dimensional compact Riemannian manifold $M$ is considered.

In fact, in \cite{Guil:85} one proves that as a consequence of the good function-theoretic properties of
$\zeta_x$ for each such operator the Weyl's formula of the asymptotic distribution of the
spectral values \cite{Weyl:12} can be seen to hold. Thus, in particular
the condition $\mu_k(x)={\mathbf O}(1/k)$ is then fulfilled automatically and does not appear as an
independent condition any longer.

But then, upon combining formula \eqref{propconn}
with a method \cite[{\sc Theorem 7.4,\,7.5}]{Guil:85} (or see \cite{Wodz:87})
of expressing the residue in terms of
the principal symbol of the classical pseudodifferential operator in question, one finally
will arrive at Connes' trace theorem.

\section{The Connes' trace theorem and its application}\label{2}
In the following we are going to comment on the way along to Connes' trace theorem
in a more detailed manner and will give some indications on applications of this formula as to classical
Yang-Mills theory.

\subsection{Preliminaries}\label{2.1}
\subsubsection{Basic facts about pseudodifferential operators}\label{2.1.1}
Let $\Omega$ be an open set in ${\mathbb{R}}^n$, and let $C^\infty_0(\Omega)$ be the
space of smooth functions with compact support inside $\Omega$.
\begin{defi}\label{symb}
Let $p\in C^\infty(\Omega\times{\mathbb{R}})$. $p$ is called a symbol of order
(at most) $m\in {\mathbb{R}}$,
if it satisfies the estimates
\begin{equation}\label{sm}
|\partial^\alpha_\xi\partial^\beta_x p(x,\xi)|\leq
C_{\alpha\beta K}(1+\|\xi\|)^{m-|\alpha|},~~x\in K,~~\xi\in{\mathbb{R}}^n,
\end{equation}
for any choice of multiindices $\alpha$, $\beta$ and compact $K\subset\Omega$.
The space of the symbols of order $m$ is denoted by $S^m(\Omega\times{\mathbb{R}}^n)$
or simply $S^m$.
\end{defi}
Note that our definition corresponds to the special case with
$\varrho=1$ and $\delta=0$ of a more general class of symbols as
considered e.g.~in \cite[Definition 1.1.]{shub-1},
to which and to \cite{ego1,ego} the reader might refer also for other details on
pseudodifferential operators.\footnote{Relating notions, conventions and
terminology, we do not follow the usage of \cite{shub-1} into any detail,
but instead join some slightly simplified conventions and notations which
are suitable for our purposes and which we borrowed from some
survey lectures of E.\,Zeidler \cite{Zeid:94}, and which are the same as in
\cite{egshu} and \cite[10.4., especially \S\,10.4.7.]{Zeid:95}.
}
It is obvious that $S^m\subset S^k$ for $m\leq k$. For $p\in S^m$, let
$p(x,D)$ denote the operator
\begin{equation}\label{pxd}
(p(x,D)u)(x)=(2\pi)^{-n/2}\int p(x,\xi)e^{i\langle x,\xi\rangle}\hat{u}(\xi)d\xi.
\end{equation}
\begin{equation}\label{four}
\hat{u}(\xi)=(2\pi)^{-n/2}\int e^{-i\langle x,\xi\rangle}u(x)dx
\end{equation}
is the Fourier transform of $u$. Note that different $p,p'\in S^m$ may lead
to the same operator, $p(x,D)=p'(x,D)$.

\begin{defi}\label{psido}
A pseudodifferential operator ($\psi$DO) of order (at most) $m$ is an
operator of the form
\begin{equation}\label{pdo}
P=p(x,D),
\end{equation}
where $p\in S^m$
.
The class of $\psi$DO's of order $m$ is denoted by $L^m$.
\end{defi}
The mapping $S^m\longrightarrow L^m$, $p\mapsto p(x,D)$, is surjective, but
in general it will not be injective. Its kernel is contained in $S^{-\infty}=
\bigcap_{m\in{\mathbb R}} S^{m}$. The $\psi$DO's  corresponding to $S^{-\infty}$ 
form the space $L^{-\infty}$ of smoothing operators\footnote{For a more precise 
argumentation the reader is referred to \cite[\S 3.2,\,Definition 3.3 and {\em{Remark}}]{shub-1}, 
and where also the notion of `properly supported $\psi$DO' comes into play, but which for simplicity 
will not be considered explicitely in this paper.}\label{class1}.
The principal symbol $\sigma_m(P)$ of a $\psi$DO $P$ of order $m$
with symbol $p\in S^m$ is the class
of $p$ in $S^m/S^{m-1}$.\\
\begin{defi}\label{class}
$p\in S^m$ is called classical, if it has an ``asymptotic expansion''
\begin{equation}\label{asy}
p\sim\sum_{j=0}^{\infty}p_{m-j},
\end{equation}
i.e. $p_{m-j}\in S^{m-j}$ and
\begin{equation}\label{asy1}
p-\sum_{j=0}^{N-1}p_{m-j}\in S^{m-N},~~\forall N,
\end{equation}
and if $p_{m-j}$ is positive homogeneous in $\xi$ ``away from $0$'', i.e.
\begin{equation}\label{hom}
p_{m-j}(x,t\xi)=t^{m-j}p_{m-j}(x,\xi),~~\|\xi\|\geq 1,~~t\geq 1.
\end{equation}
A $\psi$DO is said to be classical if its symbol is classical. The spaces of
classical symbols and $\psi$DO´s are denoted by $S^m_{cl}$ and $L^m_{cl}$
respectively.
\end{defi}
Let $p^0_{m-j}(x,\xi)$ be homogeneous functions in $\xi$ on $\Omega\times
({\mathbb{R}}^n\setminus\{0\})$ coinciding with $p_{m-j}$ for $\|\xi\|\geq 1$.
These functions are uniquely determined, and one writes also
\begin{equation}\label{asy2}
p\sim\sum_{j=0}^\infty p^0_{m-j}
\end{equation}
instead of (\ref{asy}).
The principal symbol of a classical $\psi$DO can
be identified with the
leading term $p^0_m$ in the asymptotic expansion (\ref{asy2}).
\begin{theorem}\label{diff}
Let $F:\Omega'\longrightarrow\Omega$ be a diffeomorphism of domains in ${\mathbb{R}}^n$.\\
Then to every $\psi$DO $P$ on $\Omega$ with symbol $p\in S^m(\Omega\times{\mathbb{R}}^n)$
corresponds a $\psi$DO $P'$ on $\Omega'$ with symbol $p'\in
S^m(\Omega'\times{\mathbb{R}}^n)$ such that:\\
\begin{equation}\label{corr}
F^*(Pu)=P'(F^*(u)),~~u\in C_0^\infty(\Omega),~~F^*-\mbox{pull-back},
\end{equation}
\begin{equation}\label{hsinv}
p'(x,\xi)-p(F(x),(^tF'(x))^{-1}\xi)\in S^{m-1}(\Omega'\times{\mathbb{R}}^n).
\end{equation}
If $P$ is a classical $\psi$DO then so is $P'$.
\end{theorem}
The theorem makes it possible to define $\psi$DO's on manifolds.
Let $M$ be a paracompact smooth manifold, and consider an operator
$A:C^\infty_0(M)\longrightarrow C^\infty(M)$. If $\Omega$ is some coordinate neighborhood
of $M$, there are a natural extension map $i_\Omega:C^\infty_0(\Omega)\longrightarrow
C^\infty_0(M)$ and a natural restriction map $p_\Omega:C^\infty(M)\longrightarrow
C^\infty(\Omega)$. $A$ is called $\psi$DO of order $m$ if all the local
restrictions $A_\Omega:=p_\Omega\circ A\circ i_\Omega:C^\infty_0(\Omega)\longrightarrow
C^\infty(\Omega)$ are $\psi$DO of order $m$. By Theorem \ref{diff}, this is a
good definition, and also classical $\psi$DO can be defined in this manner.
Moreover, equation (\ref{hsinv}) says that the principal symbol has an
invariant meaning as a function on the cotangent bundle $T^*M$.\\
On the other hand, $\psi$DO on a manifold can be constructed by gluing:
Let $\bigcup_j\Omega_j=M$ be a locally finite covering of $M$ by coordinate
neighbourhoods, and let $A_j$ be $\psi$DO's of order $m$ on $\Omega_j$.
Furthermore, let $\sum_j\psi_j=1$ be a partition of unity subordinate
to the given covering, and let $\phi_j\in C^\infty_0(\Omega_j)$ with
$\phi_j|_{supp~\psi_j}=1$.  Then $A:=\sum_j\phi_j\circ A_j\circ\psi_j$ ($\phi_j,~\psi_j$
considered as multiplication
operators) is a
$\psi$DO of order
$m$ on $M$ whose restrictions $A_{\Omega_j}$ coincide with $A_j$.
\vspace{.3cm}

$\psi$DO's acting on sections of vector bundles are defined with appropriate
modifications: They are glued from local $\psi$DO's which are defined using
matrices of symbols. The principal symbol is then a function on $T^*M$ with
values in the endomorphisms of $E$, i.e. a section of the bundle
$\pi^*(End(E))$,
where $\pi:T^*M\longrightarrow M$ is the projection of the cotangent bundle, and $End(E)$
is the bundle of endomorphisms of $E$.
\vspace{.3cm}

$\psi$DO's are operators from $C^\infty_0(M)$ to $C^\infty(M)$.
$\psi$DO's of order $m$ can be extended to bounded linear operators
$H^s(M)\longrightarrow H^{s-m}(M)$, $s\in{\mathbb{R}}$ (Sobolev spaces).
Notice that, by the Sobolev embedding theorems, every $\psi$DO of order
$\leq 0$, $H^s\longrightarrow H^{s-m}$,
can be considered as an operator $H^s\longrightarrow H^s$. In particular, taking the
case $s=0$, every $\psi$DO of order $\leq 0$ may be considered as an operator
$L^2\longrightarrow L^2$. For the case of manifolds, a Riemannian metric is used
in the definition of the $L^2$ scalar products, for vector bundles in addition
a fibre metric. $L^2(M,E)$ denotes the corresponding space of $L^2$ sections.
We will need the following list of facts (for some terminology and
the corresponding generalities see \cite[Definition 3.1.,\,24.3]{shub-1} and \cite[23.26.12.]{dieu7} e.g.):\\
1. The product (which exists, if at least one of the factors is ``properly
supported'') of two $\psi$DO's of orders $m$, $m'$ is a
$\psi$DO of order $m+m'$.\\
2. The principal symbol of the product of two $\psi$DO's is the product
of the principal symbols of the factors.\\
3. A $\psi$DO of order $\leq 0$ is bounded. For order $<0$ it is compact.\\
4. A $\psi$DO of order less than $-n$ on a manifold of dimension $n$ is trace
class.\\
5. If $A$ is a $\psi$DO on a manifold, and if $\phi_j$ and $\psi_j$ are as
above, then $A$ may be written
\[A=\sum_j\psi_jA\phi_j+A'\]
with $A'\in L^{-\infty}$ (smoothing operator).

\begin{remark}\label{guilrem.1}
Note that the classical $\psi$DO's form an algebra which is an example of a more
abstract object which usually is referred to as Weyl algebra.
According to \cite{Guil:85}, it is a Weyl algebra corresponding to the
symplectic cone $Y=T^*M\setminus\{0\}$ ($\{0\}$ the zero section), with its standard symplectic form
$\omega$ and ${\mathbb{R}}^+$-action $\rho_t(x,\xi)=(x,t\xi)$. That is, $Y$ is an
${\mathbb{R}}^+$-principal bundle such that $\rho_t^*\omega=t\omega$. The
properties
listed above, however, are only part of the conditions assumed in
\cite[2., A.1.-E.]{Guil:85}.
\end{remark}

\subsubsection{Definition of the Wodzicki residue}\label{2.1.2}
There are at least two equivalent definitions of the Wodzicki residue:
As a residue of a certain $\zeta$-function and as an integral of a certain local
density \cite{Wodz:87}, \cite{kass.5}.
 We take as starting point the second definition
which can be used most directly for writing classical gauge field Lagrangians.
The first definition will show up in the second proof of Connes´ theorem.
\begin{defi}
Let $M$ be an $n$-dimensional compact Riemannian manifold. Let $T$ be a
classical pseudodifferential operator of order $-n$ acting on sections of a complex
vector bundle $E\longrightarrow M$. The Wodzicki residue of $T$ is defined by
\begin{equation}\label{wod}
Res_W(T)=\frac{1}{n(2\pi)^n}\int_{S^*M}{\tr}_E\sigma_{-n}(T)\mu,
\end{equation}
where $\sigma_{-n}(T)$ is the principal symbol of $T$, $S^*M$ is the cosphere
bundle $\{\xi\in T^*M: \|\xi\|_g=1\}$ and $\mu$
is the volume element defined by a multiple of the canonical contact form on $T^*M$.
${\tr}_E$ is the natural pointwise trace on
$\pi^*(End(E))$.
\end{defi}
The form $\mu$ is defined as
$\mu=\frac{(-1)^{\frac{n(n+1)}{2}}}{(n-1)!}\alpha\wedge(d\alpha)^{\wedge(n-1)}$, where $\alpha$ is the canonical
1-form on $T^*M$, $\alpha=\sum_i\xi_idx^i$ in local coordinates.

Also $Res_W$ is defined for classical $\psi$DO of any order, using the
same formula with $p_{-n}$ instead of $\sigma_{-n}$ for integer order
$m\geq -n$, and putting $Res_W=0$ else.

It should be noted that the Wodzicki residue can be defined without
using the Riemannian structure \cite{Wodz:87}: One starts defining for a
$\psi$DO $T$ on a chart domain in ${\mathbb{R}}^n$ a matrix-valued local density
\[res_x(T)=(\int_{\|\xi\|=1}p_{-n}(x,\xi)|\bar{d}\xi|)~|dx|,\]
where $\bar{d}\xi=\sum_i(-1)^i\xi_id\xi_1\wedge\cdots\wedge\hat{d\xi_i}\wedge\cdots
\wedge\xi_n$ is the normalized volume form on the standard Euclidean sphere
$\|\xi\|=1$ and $dx$ is the standard volume form in the chart coordinates.
Note that $\bar{d}\xi dx=\mu$.
Then one shows that this has good functorial properties, i.e. is indeed a
density (an absolute value of an $n$-form) on $M$
, and defines
\[Res_W(T)=\frac{1}{n(2\pi)^n}\int_M {\mathrm {tr}}~res_x(T).\]
Due to the homogeneity property of $p_{-n}(x,\xi)$ (using the Euler formula),
$p_{-n}(x,\xi)\bar{d}\xi$ is a closed form, thus $\|\xi\|=1$ can be replaced
by any homologous $n-1$-surface, in particular by any sphere $\|\xi\|_g=1$
with respect to a chosen Riemannian metric on $M$. This leads to formula
(\ref{wod}) used above. Thus, $Res_W(T)$ does not depend on the choice of
the Riemannian metric defining the cosphere bundle. It may, however, depend
on the metric through a metric-dependence of $T$.
\\[.2cm]
\begin{remark}\label{guilrem.2}
 The residue $Res_W$ defined above coincides, up to a universal factor
which depends only on $\dim(M)$, with the residue defined in \cite[\sc{Definition
6.1}]{Guil:85}.\end{remark}
Properties of the Wodzicki residue (see \cite{Wodz:87,kass.5} and \cite[\sc{Proposition
6.1}]{Guil:85}):\\
1. $Res_W$ is a linear (in general not positive) functional on classical
$\psi$DO's.\\
2. $Res_W$ is a trace on the algebra of classical $\psi$DO's.\\
3. It is the only trace if $M$ is connected, $\dim(M)>1$.\\
4. $Res_W$ vanishes on operators of order $<-n$ or noninteger.

\subsection{Connes' trace theorem}\label{2.2}
\subsubsection{Formulation of Connes' trace theorem}\label{2.2.1}
We are now ready to formulate the famous trace theorem \cite{Conn:88}.
\begin{theorem}\label{trtheo}
Let $M$ be a compact Riemannian manifold of dimension $\dim(M)=n$, let $E\longrightarrow M$
be a complex vector bundle over $M$, and let $T$ be a classical pseudodifferential
operator of order $-n$ on $\Gamma(E)$. Then
\begin{itemize}
\item[(i)]\label{adi}
The extension of $T$ to the Hilbert space ${\mathcal H}=L^2(M,E)$ belongs to the ideal
$L^{1,\infty}({\mathcal H})$.
\item[(ii)]\label{adii}
The Dixmier trace ${\Tr}_\omega(T)$ coincides with the Wodzicki residue,
\begin{equation}\label{trfor}
{\Tr}_\omega(T)=Res_W(T)=\frac{1}{n(2\pi)^n}\int_{S^*M}{\tr}_E\sigma_{-n}(T)\mu.
\end{equation}
\end{itemize}
As a consequence, ${\Tr}_\omega(T)$ does not depend on the choice of the functional
$\omega$ in this case.
\end{theorem}
The following two parts will be devoted to proofs of this theorem exclusively. Two
variants of proving will be presented\,:

In the first variant we are following roughly the line of the original arguments
given in \cite{Conn:88}, but see also \cite{land} and \cite{GrVa:93} for some
 details\,\footnote{We are very indebted to B.\,Crell, Leipzig, who kindly placed to our disposal
his manuscript \cite{cre} and the reading of which was strongly facilitating
our understanding of some of the peculiarities of Connes' approach towards formula \eqref{trfor}.}, and the
special case of scalar operators is dealt with, essentially. Thereby, to keep
short, in some parts the
proof will be left a bit sketchy. However, in any case it will be at worst
detailed enough to convince the reader
of the validity of Connes' trace theorem for the example of the scalar operator
$(1+\Delta)^{-n/2}$ on special
compact manifold like the $n$-torus ${\mathbb{T}}^n$ or the $n$-sphere ${\mathbb{S}}^n$, respectively
($\Delta$ is the Laplacian there).

The second variant of proving will be based on an application of Corollary
\ref{connes1} and
formula \eqref{propconn}, together with some of the knowledge gained while
proving Connes' theorem
in one of the above mentioned special cases which were completely treated
in course of the first variant of the proof. Thereby, according to the arguments found in \cite{Guil:85},
we firstly learn that
\eqref{propconn} gets applicable, and secondly see that a complete proof only requires to consider
this formula explicitely for a non-trivial example (i.e.~one with non-vanishing
Dixmier-trace).
We emphasize that it is due to the pecularity of this second line of argumentation that along with a special case
then validity of the theorem in its
full generality\,--\,not only for scalar operators\,--\,can be concluded.

\subsubsection{On the proof of Connes' trace theorem}\label{2.2.2}
The idea is to see first that the theorem is true if it is true on one
manifold and then to prove it on a manifold one likes, e. g. ${\mathbb{T}}^n$ or ${\mathbb{S}}^n$.

First, the theorem is true on a manifold $M$ globally iff it is true locally.
This is due to property 5. of $\psi$DO's given above and the fact that
smoothing operators are in the kernels of both $Res_W$ and ${\Tr}_\omega$.
Now one can transport the local situation, using a local diffeomorphism, to
a local piece of another manifold $M'$. Both sides of the desired equation
do not change under this
transport. Using now again the above local-global argument, we can think
of this local operator as part of a global operator on $M'$ (gluing by means
of a partition of unity). Thus, if the theorem is true on $M'$, it must
also be true on $M$, otherwise we would have a contradiction.

Let us prove point (i) of the theorem for scalar operators on ${\mathbb{T}}^n$.
First we show $T\in L^{1,\infty}$ for any $\psi$DO of order $-n$ on ${\mathbb{T}}^n$.
The Laplacian $\Delta$ (with respect to the standard flat metric on ${\mathbb{T}}^n$)
is a differential operator of order 2, therefore $(1+\Delta)^{-n/2}$ is
a $\psi$DO of order $-n$, and $T$ can be written in the form
$T=S(1+\Delta)^{-n/2}$, where $S$ is a $\psi$DO of order 0, therefore bounded.
Since $L^{1,\infty}$ is an ideal, it is sufficient to see
$(1+\Delta)^{-n/2}\in L^{1,\infty}$.

For the proof we need yet a little result from the general theory of
compact operators.
Suppose $x\geq {\mathbf 0}$ is compact but not of finite rank. Let
$\lambda_1>\lambda_2>\ldots >0$ be the ordered sequence of the non-zero eigenvalues of $x$,
with multiplicity $m_k$ for $\lambda_k $. Then, for each integer $t\in [0,m_{k+1} ]$, $k>2$,
let us consider
\begin{subequations}\label{folge}
\begin{equation}\label{folge.1}
\gamma_{\{\sum_{j\leq k} m_j +t\}}(x)  =
\frac{\sum_{j\leq k}\lambda_j \, m_j +\lambda_{k+1} \,t}{\log \{\sum_{j\leq k} m_j +
t\}}\,,
\end{equation}
which yields all terms $\gamma_n(x) $ of the sequence \eqref{logdiv.1} with
$\sum_{j\leq k} m_j \leq n\leq \sum_{j\leq k+1} m_j $. From \eqref{folge.1} with the help of the properties of the
logarithm one then easily infers that for the mentioned $t$'s the following estimate holds\,:
\begin{equation}\label{folge.2}
 c_{k+1}^{-1}\,\gamma_{\{\sum_{j\leq k} m_j \}}(x) \leq\gamma_{\{\sum_{j\leq k} m_j +t\}}(x)
\leq  c_{k+1}\,\gamma_{\{\sum_{j\leq k+1} m_j \}}(x) \,,
\end{equation}
\end{subequations}
with $c_{k+1}=1+\{{\log (1+(m_{k+1} /\sum_{j\leq k} m_j ))}/{\log \sum_{j\leq k} m_j }\}$.
In view of the structure of the latter coefficients from \eqref{folge.2} then
the following and often useful auxiliary criterion can be seen to hold.
\begin{lemma}\label{idcrit}
Let $x\in {\mathsf {\mathcal C}B}({\mathcal H})_+
\backslash {\mathsf {\mathcal F}B}({\mathcal H})$.
Suppose $\lim_{k\to\infty} \gamma_{\{\sum_{j\leq k} m_j \}}(x) $ exists. If the sequence
$\{m_{k+1} /\sum_{j\leq k} m_j :\,k\in {\mathbb N}\}$ is bounded, then also  $\lim_{n\to\infty} \gamma_n(x) $ exists.
\end{lemma}

Now we are ready to start our considerations around $({\mathbf 1}+\Delta)^{-n/2}$.
The spectrum of the Laplacian $\Delta=-\sum_{i=1}^n\partial_i^2$ on
${\mathbb{T}}^n={\mathbb{R}}^n/2\pi{\mathbb{Z}}^n$
is pure point, consisting of the values $\sum_ik_i^2$, $k_i\in{\mathbb{Z}}$.
The corresponding eigenfunctions are $e^{ikx}$, $k\in{\mathbb{Z}}^n$, $x\in {\mathbb{R}}^n$.
The multiplicity of an eigenvalue $\lambda$ of $\Delta$ is
$m(\lambda)=\#\{k\in{\mathbb{Z}}^n|\sum_ik_i^2=\lambda\}$. From this follow analogous facts
for the operator $(1+\Delta)^{-n/2}$. Let $m_k$ be the multiplicity of the $k$-th eigenvalue of the latter.
Let
$$
{\tilde{\gamma}}_R((1+\Delta)^{-n/2})=\frac{\sum_{1+\|k\|^2\leq R^2}(1+\|k\|^2)^{-n/2}}{\log
N'_R},
$$
where $N'_R$ is the number of lattice points in ${\mathbb{Z}}^n$ with $1+\|k\|^2\leq R^2$.
By construction it is easily seen that convergence of
$\{{\tilde{\gamma}}_R((1+\Delta)^{-n/2}):R\in {\mathbb R}_+\backslash\{0\}\}$
as $R\to\infty$
implies the limit $\lim_{k\to\infty} \gamma_{\{\sum_{j\leq k} m_j \}}((1+\Delta)^{-n/2}) $ of the considered
subsequence
of the sequence \eqref{logdiv.1} to exist (in which case then both
limits have the same value). It is not hard to see that for geometrical reasons with the above
multiplicities also the other condition in the hypotheses of Lemma \ref{idcrit} is
fulfilled; this e.g.~can be concluded as a by-result from our estimates given below and relating
the asymptotic behavior of the ratio between the surface of an $n$-sphere to the volume of the $n$-ball
of the same radius $R$ within ${\mathbb R}^n$. Hence, in view of Lemma \ref{idcrit} the conclusion is that
if the limit
\[\lim_{R\to\infty} {\tilde{\gamma}}_R((1+\Delta)^{-n/2})=
\lim_{R\to\infty}\frac{\sum_{1+\|k\|^2\leq R^2}(1+\|k\|^2)^{-n/2}}{\log
N'_R}\]
can be shown to exist, then by Theorem \ref{ex}\,\eqref{ex.2} it has to equal
${\Tr}_\omega{(1+\Delta)^{-n/2}}$ (independent of
$\omega$). Also, it is not hard to see that the latter limit exists if
\[\lim_{R\to\infty}\frac{\sum_{\|k\|\leq R}\|k\|^{-n}}{\log N_R}\]
exists, where $N_R$ is the number of lattice points with $\|k\|\leq R$,
in which case then both limits yield the
same value. We prove that the latter limit exists, computing its value.
It is well known \cite{walf} that
\[N_R=V_R+O(R^{\frac{n-1}{2}}),\]
where $V_R=\frac{\Omega_n}{n}R^n$ (volume of the ball of radius $R$ in ${\mathbb{R}}^n$),
$\Omega_n=\frac{2\pi^{n/2}}{\Gamma(n/2)}$ (area of the sphere $S^{n-1}$).
Neglecting terms of lower order in $R$, we have
\[N_R=V_R+\ldots = \frac{\Omega_n}{n}R^n+\ldots.\]
In order to determine $\sum_{\|k\|\leq R}\|k\|^{-n}$ for large $R$, we first count
the number of lattice points in a spherical shell between $R$ and $R+dR$,
\[N_{R+dR}-N_R=V_{R+dR}-V_R+\ldots=\Omega_nR^{n-1}dR+\ldots.\]
Integrating this, we obtain asymptotically for large $R$
\[\sum_{\|k\|\leq R}\|k\|^{-n}=\Omega_n\int_1^Rr^{-n}r^{n-1}dr+\ldots
=\Omega_n\log R+\ldots.\]
Together with $\log N_R=n\log R+\log\Omega_n-\log n+\ldots$ this leads to
\[\lim_{R\to\infty}\frac{\sum_{\|k\|\leq R}\|k\|^{-n}}{\log N_R}=\frac{\Omega_n}
{n},\]
i.e.
\begin{equation}
{\Tr}_\omega{(1+\Delta)^{-n/2}}=\frac{\Omega_n}{n}.
\end{equation}
It is much easier to determine the Wodzicki residue of $(1+\Delta)^{-n/2}$\,:
the principal symbol of $(1+\Delta)^{-n/2}$ is
$\sigma_{-n}((1+\Delta)^{-n/2})(x,\xi)=\|\xi\|^{-n}$, where $\|.\|$ denotes the
standard euclidean metric on ${\mathbb{R}}^n$. Therefore,
\[Res_W((1+\Delta)^{-n/2})=\frac{1}{n(2\pi)^n}\int_{S^*{\mathbb{T}}^n}\bar{d}\xi dx
=\frac{1}{n(2\pi)^n}\Omega_n\int_{{\mathbb{T}}^n}dx=
\]
\[
=\frac{1}{n(2\pi)^n}\Omega_n(2\pi)^n
=\frac{\Omega_n}{n},\]
coinciding with the result for the Dixmier trace. Thus, the theorem is already
proved for a special operator on ${\mathbb{T}}^n$.

To prove point (ii) of the theorem, we start with some general remarks about
${\Tr}_\omega$. It is a positive linear functional on the space $L^{-n}$ of
$\psi$DO's of order $-n$ with $L^{-n-1}\subset \ker {\Tr}_\omega$, because elements
of $L^{-n-1}$ are trace class (see property 4. above). 
Since always $L^{-n}/L^{-n-1}\simeq S^{-n}/S^{-n-1}$ is fulfilled\footnote{cf.~second footnote on page \pageref{class1}.} it then follows 
that ${\Tr}_\omega$ may be considered as a linear
functional on $S^{-n}/S^{-n-1}$, the space of principal symbols of $\psi$DO's
of order $-n$. By restriction, it is also a linear functional on the space
of principal symbols of classical $\psi$DO's of order $-n$. On the other hand,
this latter space and the space $C^\infty(S^*M)$ coincide, since every element
of $C^\infty(S^*M)$ by homogeneity defines a classical principal symbol,
(\cite{dieu7}, 23.29.11.). Thus, we end up with a linear functional
on $C^\infty(S^*M)$. It follows from symbol calculus that this functional is
positive (see \cite{cre}).
Thus, we have a positive distribution, which is always given by a positive
measure on $S^*M$ (\cite{dieu3}, 17.6.2).

Since an isometry of $M$ gives rise to a unitary transformation of $L^2(M,v_g)$,
and the spectrum of an operator does not change under unitary transformations,
the Dixmier trace is invariant under isometries. Therefore, the corresponding
measure on $S^*M$ is invariant under isometries.

Considering now the case
$M={\mathbb{S}}^n$, the standard $n$-sphere with the metric induced from the euclidean
metric on ${\mathbb{R}}^{n+1}$, the group of isometries is $SO(n+1)$, and $S^*M$ is a
homogeneous space under the induced action of $SO(n+1)$. It is easy to see
that the volume form of the induced Riemannian metric on $S^*{\mathbb{S}}^n$ is invariant
under the action of $SO(n+1)$. Uniqueness of the invariant measure on a
homogeneous space shows that the positive measure corresponding to the
Dixmier trace must be proportional to the measure given by this volume form
$v_g$,
\[{\Tr}_\omega(T)=const. \int_{S^*M}\sigma_{-n}(T)v_g.\]
It is an easy exercise to show that $v_g$ coincides in this case with the form
$\mu$ defined above. Moreover, the constant, which neither depends on the
operator nor on the Riemannian manifold, is determined by the example of the
torus.
\begin{remark}\label{res.4}
Note that it follows from Corollary \ref{connes1} and \ref{trtheo} that the Wodzicki
residue in some cases coincides with a residue of the zeta function
$\zeta_T(z)=\tr T^z$ (see also below).
\end{remark}

\subsubsection{An alternative proof of the Connes' trace formula}\label{2.2.3}
In this part another derivation of the trace formula \eqref{trfor} will be
presented. Thereby, the line will be to make the abstract formula
\eqref{propconn} directly accessible for Connes' trace formula in the
special case of a classical pseudodifferential operator $T$ of order $-n$. Thereby, by some continuity argument,
it suffices if the case of positive-definite operators of that kind can be dealt with.

According to the hypotheses in Corollary \ref{connes1} under which
\eqref{propconn} is supposed to hold, for this aim it is sufficient to know that
the `right' asymptotic behavior of the spectral values of $T$ holds, and in which case then
the extensions of the $\zeta_T$-function of \eqref{wiener.2} have to be analyzed.

Both exercises can be achieved at once and almost without proof
by means of V.\,Guillemin's methods given in \cite{Guil:85}. In demonstrating this way
towards formula \eqref{trfor} we finally will end up with an
alternative proof of Theorem \ref{trtheo}.

Before doing this, we recall the special settings corresponding to
the assumptions of Theorem \ref{trtheo}.
In line with these and in accordance with {\em Remark} \ref{guilrem.1} the
Weyl algebra ${\mathcal W}$ of all classical $\psi$DO's corresponding to
the symplectic cone $Y=T^*M\backslash \{0\}$ will be considered.
We then have the Hilbert space ${\mathcal H}=L^2(M,E)$, which is the completion of the sections $\Gamma(E)$ under
a scalar product descending from a symplectic volume element $\mu$ on $Y$ and a
fibre metric on $E$. If these are fixed,
each classical $\psi$DO $T$ which is at most of order $0$ corresponds to a
bounded linear operator, and can be
identified with its unique bounded linear extension $x=T$ from sections
$\Gamma(E)$ onto the whole $L^2(M,E)$.
Accordingly, in such case we use the same notation $T$ for both the
$\psi$DO and its unique bounded
linear extension on all of $L^2(M,E)$.

Now, let $x$ be a bounded linear operator on ${\mathcal H}$. Then, in the special case that $x$ is
positive-definite, there exists in a unique way the inverse
of $x$ on ${\mathcal H}$, that is, a
densely defined, positive-definite self-adjoint linear operator $x^{-1}$ on
${\mathcal H}$, with $x^{-1}\, x={\mathbf 1}$ and $x\, x^{-1}\subset {\mathbf 1}$.
In this case we then define $P_x$ to be the $n$-th (positive) root of this inverse, $P_x=\sqrt[n]{x^{-1}}$.

Especially, if $T\in L^{-n}_{cl}$($={\mathcal W}^{-n}$ in the terminology of \cite{Guil:85}) is supposed to be
positive-definite, by compactness and in view of
the definition of\,--\,and the properties coming along with\,--\,the term
ellipticity, for $x=T$ the operator $x^{-1}$, and thus also $P_x$, is positive-definite, self-adjoint, elliptic
and of order one.

Now, from \cite[(1.1)]{Guil:85} one in particular learns that for the asymptotic growth of the
singular values of a positive-definite, self-adjoint elliptic differential operator $P$ of order one
the Weyl's formula \cite{Weyl:12} holds. This equivalently says that the singular values behave like
$\mu_k(P)\sim l \sqrt[n]{k}$, with some constant $l$, see \cite[(13.18) and Proposition 13.1]{shub-1}.
Hence, in the above-mentioned special case $P=P_x$ this asymptotic law amounts to
$\mu_k(T)=\mu_k(x)\sim L \cdot k^{-1}$, with some constant $L\geq 0$, for each positive-definite
$T\in L^{-n}_{cl}(={\mathcal W}^{-n})$. Especially, in line with Lemma \ref{ike4}
we then have $T\in L^{1,\infty}({\mathcal H})$, for each such operator.
Since $L^{1,\infty}({\mathcal H})$ is an ideal from this especially $L^{-n}_{cl}\subset L^{1,\infty}({\mathcal H})$
is seen, which demonstrates that Theorem \ref{trtheo}\,(i) can be equivalently followed also from the main
result of \cite{Guil:85}.

On the other hand, owing to Weyl's asymptotic law, the applicability of
formula \eqref{propconn} now will rely on the extension properties of $\zeta_x$ from the half-plane $\Re z >1$
onto ${\mathbb C}$ exclusively. This matter we are
going to discuss now.

We start with some preliminary considerations about various existing definitions relating to
$\zeta$-functions which can be associated to some positive operator.\footnote{We have to thank H.\,Upmeier,
Marburg, who outlined to us some of the relevant details.}

Firstly, in accordance with the above and Corollary \ref{connes1} the
$\zeta_x$-function as given in \eqref{wiener.2} for compact positive
$x\in  L^{-n}_{cl}\subset L^{1,\infty}({\mathcal H})$ is holomorphic in the half-plane $\Re z>1$.
Thus for given fixed $n\in {\mathbb N}$ upon defining $\zeta^x(w)=\zeta_x(z)$ at $w=n(1-z)$, $\Re z>1$,
one gets another complex function $w\mapsto\zeta^x(w)(=\zeta_n^x(w))$ which is holomorphic in the
half-plane $\Re w<0$.
Recall that in view of the arguments given in context of \eqref{wiener.2} the operator family
$\Re z>1:\,z\mapsto\,x^z$ consists of trace-class operators. But then, by functional calculus
this operator family may be equivalently re-defined at each $w\in {\mathbb C}$ with
$\Re w<0$ and $z=(1-w/n)$ as $\Re w<0,\,w\,\mapsto\,x P_x^w$.
Hence, if another $\zeta$-function
\begin{equation}\label{zetaguil}
\forall\,w\in {\mathbb C},\Re w<0:\ \zeta(x,P)(w)=\tr x P^w
\end{equation}
is defined for bounded-invertible,
positive-definite self-adjoint linear operator $P$, and bounded
$x\geq {\mathbf 0}$ such that
$ x P^w\in L^1({\mathcal H})$ for all $w$ with $\Re w<0$, in view of the above in the
special case of $P=P_x$ we may summarize as follows.
\begin{lemma}\label{guil.1}
Suppose $x\in  L^{-n}_{cl}$ to be positive-definite.
Then, in the special case of
$P=P_x$ the complex function $w\mapsto\zeta(x,P)(w)$ is holomorphic in the
half-plane $\Re w<0$, and there $\zeta(x,P)(w)=\zeta_x(z)$ is fulfilled,
at $z=(1-w/n)$.
\end{lemma}
Now, from \cite[{\sc Theorem 7.4}]{Guil:85} one also knows that the nuclear dimension
of ${\mathcal W}$ is $n$ (the conclusion of Theorem \ref{trtheo}\,(i) being
in accordance with this), where $n$ is the dimension of the basic manifold $M$.
Hence and especially, if $x=T\in L^{-n}$ is a classical $\psi$DO
($\in{\mathcal W}^{-n}$ in the terminology of \cite{Guil:85}),
then for each positive-definite, self-adjoint {\em elliptic}\, operator $P\in {\mathcal W}^1$
(that is, $P$ is of order one, and among other facts, has to be bounded-invertible on ${\mathcal H}$, e.g.)
the $\zeta$-function
$w\,\mapsto\,\zeta(T,P)(w)$ of \eqref{zetaguil} may be considered. In fact, as a consequence of
positive-definiteness and ellipticity of $P$ and since $T$ is of order $-n$
(which is the negative of the nuclear dimension of
${\mathcal W}$) all the conditions under which  \eqref{zetaguil} is to hold are fulfilled,
and thus the restriction to the negative half-plane of
the trace of the operators
$T(w)$, whose operator family is given as $T(w)=T\,P^w\in{\mathcal W}^{w-n}\subset {\mathcal W}^w$
for $w\in {\mathbb C}$, makes sense. Thereby, the mentioned operator family itself is known to possess a
canonical property;
it is a so-called {\em holomorphic} family of operators
\footnote{For the precise definition and basic properties
around
{\em ellipticity}\, and {\em holomorphy} for $\psi$DO's of a given order and operator families, respectively,
we refer to \cite[{\sc Definition 2.1,\,Proposition 4.1},\,and eq.\,(3.18)]{Guil:85}}.

The latter especially means that the conditions of the hypothesis of
\cite[{\sc Theorem 7.1}]{Guil:85} are fulfilled, and then in line with the conclusion of this result
$\zeta(T,P)(w)$ has to be holomorphic in the half-plane $\Re w<0$ and has a
meromorphic extension to the whole complex
plane, and at $w=0$ has, at worst, a simple pole. Moreover, according to \cite[{\sc Theorem 7.4}]{Guil:85}
the residue of this meromorphic extension depends only on the symbol $\sigma_{-n}(T)$ of $T$, and has the
form
\begin{subequations}\label{res}
\begin{equation}\label{res.1}
{\mathfrak Res}|_{w=0}\,\zeta(T,P)=g_0\,{\mathrm Res}(\sigma_{-n}(T))\,,
\end{equation}
with a non-zero constant, $g_0\not=0$,
which depends only on the Weyl algebra ${\mathcal W}$ under consideration. Now,
remind that we are in the special
context described in {\em Remark~\ref{guilrem.1}}. But then, the
 cosphere bundle
$S^*M=\{\xi\in T^*M: \|\xi\|=1\}$ appears as the (compact) base of the
symplectic cone $Y=T^*M\backslash \{0\}$.
Denoting by $\pi':Y \rightarrow S^*M$ the projection of $Y$, one defines
for a homogeneous (of degree $-n$) $C^\infty$-section $f$ of the bundle
$\pi^*End(E)\rightarrow Y$, according to
\cite[{\sc Definition 6.1}]{Guil:85},


\begin{equation}\label{res.2}
{\mathrm Res}(f)=\int_{S^*M} {\tr}_E f\tilde{\mu}=\int_{S^*M} \mu_f\,.
\end{equation}
Here $\tilde{\mu}=\alpha\wedge\omega^{\wedge(n-1)}$ is the volume element
defined by a salar multiple of the canonical contact form on $T^*M$ (and
thus in accordance with {\em Remark \ref{guilrem.1}} one has
$\rho_t^*\,\tilde{\mu}=t^n\,\tilde{\mu}$,
$\mu_f$ is the uniquely determined
$(2n-1)$-form defined through ${\tr}_Ef\tilde{\mu}={\pi'}^*(\mu_f)$ and
${\tr}_E$ is the natural pointwise trace on $\pi^*(End(E))$.
As mentioned above there is a basic fact saying that homogeneous
$C^\infty$-sections of $\pi^*End(E)\rightarrow Y$ in our case exactly yield all the
principal symbols
to classical $\psi$DO's of order $-n$.

Thus, the above-mentioned conclusions about the possibility of a meromorphic extension
of $\zeta(T,P)$ and its singularity structure at $w=0$ apply with $P=P_T$. In view of Lemma \ref{guil.1} and
as a consequence of the just said, upon changing the complex variable $w$ into $z$ in accordance with $w=n(1-z)$
we will see that analogous facts hold in respect of $\zeta_T$ and at $z=1$, accordingly.
That is, $\zeta_T$ possesses a meromorphic extension into the whole $z$-plane,
with a simple pole at $z=1$, at worst. Having in mind this, and taking into account that from $w=n(1-z)$ a
geometric factor $1/n$ arises while passing from the
residue of the one extension at $w=0$ to the residue of the
transformed extension at $z=1$, in view of \eqref{res.1}--\eqref{res.2} we then may summarize as follows\,:
\begin{corolla}\label{guilres}
Let $T\in L_{cl}^{-n}$ be positive-definite. The holomorphic function $\zeta_T(z)$ has a
meromorphic extension from the half-plane
$\Re z>1$ into the whole complex plane with, at worst,
a simple pole at $z=1$. The residue of the extension obeys
$${\mathfrak Res}|_{z=1}\,\zeta_T(z)=(g_0/n)\int_{S^*M} {\tr}_E \sigma_{-n}(T)\mu\,,$$
with a constant
$g_0\not=0$ which does not depend on the special operator $T$.
\end{corolla}
Foremost, according to Corollary \ref{connes1} and by the above-mentioned asymptotic spectral properties
the Corollary guarantees that formula \eqref{propconn} can be
applied for positive-definite classical $\psi$DO's of order $-n$, with the result that
\begin{equation}\label{res.3}
{\Tr}_\omega(T) =(g_0/n)\int_{S^*M} {\tr}_E \sigma_{-n}(T)\mu
\end{equation}
\end{subequations}
has to be fulfilled, for each positive-definite $T$ of order $-n$.
Also, in order to fix the constant $g_0$ it obviously suffices to deal with
one particular case of such an operator. Moreover, once more again according to
the local-global and the $M$-to-$M^\prime$ arguments, which we have already
mentioned at the beginning of \ref{2.2.2} while proceeding the first
variant of the proof of Theorem \ref{trtheo}, we have to conclude that the constant $g_0$ within
\eqref{res.3} has to be the same, in each case of an $n$-dimensional compact manifold $M$.
Hence, we may content with the known result for $T=({\mathbf 1}+\Delta)^{-n/2}$ on the $n$-torus $M={\mathbb{T}}^n$.
According to our calculations therefore $g_0=1/(2\pi)^n$ has to hold, and then \eqref{res.3}
will yield that \eqref{trfor} has to be valid, for each positive-definite $T$
of order $-n$ on an arbitrary compact $n$-dimensional manifold $M$.
>From this the validity for all positive $T$ of order $-n$ can be concluded, since for fixed positive-definite
$T_0$ of order $-n$ and
each positive $T$ the family $T(\varepsilon)=T+\varepsilon\,T_0$, $\varepsilon>0$, consists of
positive-definite $\psi$DO's of order $-n$, for which according to the above the assertion of Connes formula
holds. In fact, according to Lemma \ref{lini} one knows
${\Tr}_\omega(T(\varepsilon)) = {\Tr}_\omega(T) +\varepsilon\,{\Tr}_\omega(T_0)$. On the other hand, the
map $A\mapsto \sigma_{-n}(A)$ between $\psi$DO's of order $-n$ and their principal symbols is a homomorphism,
and therefore also $\sigma_{-n}(T(\varepsilon))=\sigma_{-n}(T)+
\varepsilon\,\sigma_{-n}(T_0)$. Hence, since the expression on the right-hand
side of \eqref{res.3} obviously is a linear form with respect to the
$\sigma_{-n}(T)$-variable, the validity of
\eqref{trfor} in the general case can be obtained simply via the just mentioned linearity and
upon taking the difference between a
relation of type \eqref{res.3}, taken at one particular
$T(\varepsilon)$, for some $\varepsilon>0$, and a multiple of the
relation of type
\eqref{res.3} at $T_0$ with $\varepsilon$.

\subsection{Classical Yang-Mills actions}\label{2.3}

Here we make some remarks about the construction of the bosonic
part of classical (pure) gauge field actions
in terms of the Dixmier trace and the classical Dirac operator. This was considered in
more detail in the lectures by R. Holtkamp and K. Elsner/H. Neumann.
We will make use of the fact that the de Rham algebra of exterior forms is
isomorphic to the differential algebra $\Omega_D(C^\infty(M))$ coming from the
classical spectral triple $({\cal A}=C^\infty(M),{\cal H}=L^2(M,S),D)$, $D$ the
Dirac operator on the compact $n$-dimensional Riemannian spin manifold $M$, $S$
the spinor bundle (see the lecture by M. Frank). The representation $\pi$ of ${\cal A}$ on
${\cal H}$ is given by sending $f\in {\cal A}$ to the operator of multiplication with
the function $f$.

\subsubsection{The classical Dirac operator and integration on manifolds}\label{2.3.1}
First, we notice
\begin{satz}\label{diint}
Consider $f\in C^\infty(M)$ as left multiplication operator on $L^2(M,S)$.
Then
\begin{equation}
{\Tr}_{\omega}\left({f|D|^{-n}}\right)=\frac{1}{c(n)}\int_Mf ~v_g,
\end{equation}
where $v_g$ denotes the Riemannian volume element,
$c(n)=2^{n-[n/2]-1}\pi^{n/2}n\Gamma(n/2)$, and ${\Tr}_{\omega}$ is the Dixmier
trace with respect to any invariant mean $\omega$.
\end{satz}
\begin{proof} (see \cite{land}, p. 98)
The principal symbol of the Dirac operator is $\gamma(\xi)$ (Clifford
multiplication on spinors), thus $D$ is a first order (elliptic) $\psi$DO.
Multiplication with $f$ is a zero order operator, therefore $f|D|^{-n}$
is a $\psi$DO of order $-n$. Its principal symbol is $\sigma_{-n}(x,\xi)=
f(x)\|\xi\|^{-n}1_{2^{[n/2]}}$, where $1_{2^{[n/2]}}$ is the identity map of the
fibre $S_x$ of $S$. This principal symbol reduces on the cosphere bundle $S^*M$
to $f(x)1_{2^{[n/2]}}$. Thus, Theorem \ref{trtheo} gives
\[
{\Tr}_{\omega}\left({f|D|^{-n}}\right)=\frac{1}{n(2\pi)^n}\int_{S^*M}{\tr}_S(f(x)1_{2^{[n/2]}})dx
\bar{d}\xi
\]
\[=\frac{2^{[n/2]}}{n(2\pi)^n}\int_{S^{n-1}}\bar{d}\xi\int_M f(x)dx.\]
The area $\int_{s^{n-1}}\bar{d}\xi=\frac{2\pi^{n/2}}{\Gamma(n/2)}$ of the unit sphere
$S^{n-1}$ leads to the right factor $c(n)$.\end{proof}

Since $|D|^{-n}$ is in $L^{1,\infty}$ we can define the following inner product
on $\pi(\Omega^k{\cal A})$:
\begin{equation}\label{inn}
\langle T_1,T_2\rangle_k:={\Tr}_\omega\left({T_1^*T_2|D|^{-n}}\right).
\end{equation}
In order to really have an inner product, one needs some assumptions about
${\cal A}$ which are fulfilled in the classical case, but which also hold in more general
situations of spectral triples, see \cite{cigusc,land,vari:97}.\footnote{Thanks to J.\,V\'{a}rilly
for pointing out this fact to us.}
The orthogonal complement with respect to this inner product of the
subspace $\pi(d(J_0\cap\Omega^{k-1}))\subset \pi(\Omega^k{\cal A})$ is isomorphic
to $\Omega_D^k{\cal A}$ (both are images of surjections with the same kernel).
Thus, the inner product (\ref{inn}) can be tranported to $\Omega_D^k{\cal A}$.
\begin{satz}\label{inncl}
Under the isomorphism between $\Omega_D^k{\cal A}$ and $\Gamma(\Lambda_{{\mathbb{C}}}T^*M)$
the inner product on $\Omega_D^k{\cal A}$ is proportional
to the usual Riemannian inner product,
\begin{equation}\label{innrie}
\langle\omega_1,\omega_2\rangle_k=(-1)^k\lambda(n)\int_M\omega_1\wedge {}^*\omega_2
\end{equation}
for $\omega_i\in \Omega_D^k{\cal A}\simeq \Gamma(\Lambda_{{\mathbb{C}}}T^*M)$, where
\[\lambda(n)=\frac{2^{[n/2]+1-n}\pi^{-n/2}}{n\Gamma(n/2)}.\]
\end{satz}
\begin{proof} We refer to \cite{land}, p.~120.\end{proof}

\subsubsection{Classical gauge field actions in terms of Dixmier-trace}\label{2.3.2}
Now, in usual gauge theory, the gauge field $F$ may be interpreted as a two-form
with values in the endomorphisms of a vector bundle $E$ over $M$ (curvature of a connection).
(Such vector bundles typically arise as bundles
associated to a principal bundle with the group of inner symmetries as structure group).
The (pure) gauge field
action is then constructed by combining the scalar product on the right-hand side
of (\ref{innrie}) with a (fiberwise) product and trace of the endomorphisms,
\[
YM(\nabla)=const.\int_M tr(F\wedge *F).
\]

By (\ref{innrie}) this can be written equivalently in terms of the differential
algebra $\Omega_D^k{\cal A}$ and the scalar product there.

Moreover, it is almost obvious from the
definition of the inner product that the classical $YM$ action can be obtained
as an infimum over a ``universal'' $YM$ action defined over universal connections
(which are elements of $\Gamma(End E)\otimes_{C^\infty(M)}\Omega^2({\mathbb{C}}^\infty(M))$.
More precisely \cite{Conn:94}, \cite{land}, one shows that
\[
1\otimes \pi:{\cal E}\otimes_{\cal A}\Omega^1{\cal A}\longrightarrow
{\cal E}\otimes_{\cal A}\Omega^1_D{\cal A}
\]
(${\cal E}:=\Gamma(E)$)
gives rise to a surjection from universal connections to usual connections.
If $\theta$ is the curvature of a universal connection $\nabla_{un}$, one
defines
\[
I(\nabla_{un})={\Tr}_\omega\left(\{1\otimes \pi\}(\theta)^2|1\otimes D|^{-n}\right).
\]
and finds
\[YM(\nabla)=const.\inf\{I(\nabla_{un})|\pi(\nabla_{un})=\nabla\}.\]
Thus the classical Yang-Mills action can be entirely written in terms
of objects which have a straightforward generalization to the noncommutative
situation.
\newpage
{\bf Acknowledgements}
\vspace{.3cm}

We thank Harald Upmeier (Marburg) and Wend Werner (Paderborn)
for inviting us to give this lecture. Special thanks go to Harald Upmeier for insisting on
the use of Guillemin's results for a proof of Connes' formula. In particular,
we are grateful to him for hints relating $\zeta$-functions, and for supplying Portenier's notes.

We are very indebted to Gianni Landi (Trieste, Italy), Claude Portenier (Marburg) and
Joseph C.\,V\'{a}rilly
(San Jos\'{e}, Costa Rica) for helpful comments, examples and
hints to the literature.

Last but not least, we are grateful to all those
collegues who helped us in
reconstructing and extracting some of the valuable informations reminiscent of the
`Oberseminar zur Mathematischen Physik'
which took place as a joint seminar  during the Academic course 1994/95 at the Mathematics and Theoretical
Physics Departments of Leipzig University.
In this respect, special thanks go to Bernd Crell, Konrad Schm\"udgen and Eberhard Zeidler
(all from Leipzig)
whose notices of their
contributed lectures to the Oberseminar together with many clarifying discussions
around the
subject proved particularly useful for us.

\bibliographystyle{abbrv}

\end{document}